\RequirePackage{ifpdf}
\ifpdf 
\documentclass[pdftex]{sigma}
\else
\documentclass{sigma}
\fi

\usepackage{slashed}
\usepackage{bm}
\usepackage{bbm}
\usepackage[mathscr]{eucal}

\newcommand{\Tr}{\hbox{Tr}}
\newcommand{\Eins}{\mathbbmss{1}}
\renewcommand{\Eins}{\mathbbmss{1}}
\newcommand{\starc}[2]{\left[{#1}\,\overset{\star}{,}\,#2\right]}
\renewcommand{\sp}{\!\cdot\!}
\renewcommand{\wp}{\!\wedge\!}
\newcommand{\m}{{\scriptscriptstyle -}} 
\newcommand{\p}{{\scriptscriptstyle +}}
\newcommand{\LCpm}{{\scriptscriptstyle \pm}}

\newcommand{\LCperp}{{\scriptscriptstyle \perp}}

\newcommand{\ud}{\mathrm{d}}

\newcommand{\E}{\mathbf{E}}
\newcommand{\B}{\mathbf{B}}
\newcommand{\D}{\mathscr{D}}

\begin{document}

\allowdisplaybreaks

\renewcommand{\thefootnote}{$\star$}

\renewcommand{\PaperNumber}{041}

\FirstPageHeading

\ShortArticleName{Strong Field, Noncommutative QED}

\ArticleName{Strong Field, Noncommutative QED\footnote{This paper is a
contribution to the Special Issue ``Noncommutative Spaces and Fields''. The
full collection is available at
\href{http://www.emis.de/journals/SIGMA/noncommutative.html}{http://www.emis.de/journals/SIGMA/noncommutative.html}}}

\Author{Anton ILDERTON, Joakim LUNDIN and Mattias MARKLUND}

\AuthorNameForHeading{A.~Ilderton, J.~Lundin and M.~Marklund}

\Address{Department of Physics, Ume\aa\ University, 901-87 Ume\aa, Sweden}
\Email{\href{mailto:anton.ilderton@physics.umu.se}{anton.ilderton@physics.umu.se}, \href{mailto:joakim.lundin@physics.umu.se}{joakim.lundin@physics.umu.se}, \\
\hspace*{13.5mm}\href{mailto:mattias.marklund@physics.umu.se}{mattias.marklund@physics.umu.se}}

\ArticleDates{Received March 23, 2010, in f\/inal form May 17, 2010;  Published online May 26, 2010}

\Abstract{We review the ef\/fects of strong background f\/ields in noncommutative QED. Beginning with the noncommutative Maxwell and Dirac equations, we describe how combined noncommutative and strong f\/ield ef\/fects modify the propagation of fermions and photons. We extend these studies beyond the case of constant backgrounds by giving a new and revealing interpretation of the photon dispersion relation. Considering scattering in background f\/ields, we then show that the noncommutative photon is primarily responsible for generating deviations from strong f\/ield QED results. Finally, we propose a new method for constructing gauge invariant variables in noncommutative QED, and use it to analyse the physics of our null background f\/ields.}

\Keywords{noncommutative QED; background f\/ields}

\Classification{81T75; 81T13}

\renewcommand{\thefootnote}{\arabic{footnote}}
\setcounter{footnote}{0}

\section{Introduction}

There has been a recent resurgence of interest in exploring the vacuum of QED using intense lasers, through `nonlinear vacuum phenomena', as the electric f\/ields available at current and near future laser facilities continue to approach the Sauter--Schwinger limit \cite{Heinzl:2008an, Marklund:2008gj}.  In this paper we investigate the ef\/fects of a strong background f\/ield on physics beyond the standard model. Our focus here is on noncommutative QED (NCQED for short), where we presuppose the existence of some space-time noncommutativity, characterised by the noncommutativity tensor $\theta^{\mu\nu}$, and investigate its consequences. For introductions covering diverse aspects of noncommutative f\/ield theory see for example \cite{Landi:1997sh, Douglas:2001ba, Szabo:2001kg, Szabo:2004ic}, and see \cite{Carroll:2001ws, Szabo:2009tn}  plus references therein for reviews of experimental signatures of noncommutativity.

Rather than considering processes in vacuum, as would be appropriate for collider based experiments, we will here be interested  in processes within a strong background f\/ield, and which typically do not occur in vacuum. Since NCQED scattering amplitudes share the same momentum conservation laws as their QED counterparts, we will therefore be examining combined noncommutative and strong f\/ield ef\/fects. The most commonly considered backgrounds in such investigations are magnetic, as they are naturally tied to the most familiar case of space-space noncommutativity, and because the strongest background f\/ields available in astrophysical contexts are magnetic (with some observations indicating that the magnetic f\/ield strength in magnetars may exceed the QED critical value at which the cyclotron energy equals the electron rest mass  \cite{Kouveliotou:1998ze, Palmer:2005mi, mag3}). The strongest terrestrial source of magnetic f\/ields, though, are in high intensity laser facilities.  Because of this, we will take our backgrounds to be plane waves, or `null f\/ields', which are commonly used to model the f\/ields of a laser. These backrounds are characterised by light-like vectors, and as a result the physics we consider will be sensitive to the components~$\theta^{+\nu}$ of the noncommutativity tensor, and hence probe {\it light-like} noncommutativity~\cite{Aharony:2000gz}.

\looseness=1
We should say a few words about light-like noncommutative theories here. In order to avoid unitarity problems~\cite{Gomis:2000zz}, theories with time-space noncommutativity must be carefully def\/ined~\cite{Bahns:2002vm}.  As for theories with space-space noncommutativity, though, the usual Feynman diagram expansion of a light-like noncommutative theory is unitary~-- provided that $\theta^{+-}=0$. From this perspective, light-like noncommutative theories are in the same class as those with space-space noncommutativity.  It therefore makes sense as a f\/irst step to follow the standard approach to space-space noncommutativity by considering a particle in a strong background f\/ield. Rather than a magnetic f\/ield, the appropriate background is a `crossed f\/ield' where the electric and magnetic f\/ields are orthogonal, and of equal magnitude (the long wavelength limit of a plane wave). The behaviour of a particle in a strong crossed f\/ield is very dif\/ferent from that in a magnetic f\/ield, though, and while an analogy can be made with the decoupling of states which leads to projection onto the lowest Landau level seen in the magnetic case, the resulting noncommutative theory seems be one of the notorious lightcone zero modes \cite{Heinzl:2009rf}. How one should proceed in this regard is an open problem, so instead we begin here with noncommutative QED as a model of physics beyond the standard model, and investigate the consequences of space-time noncommutativity.

\looseness=1
We begin with the essential properties of QED in strong background f\/ields, summarising calculations in the Furry picture, the basic lepton-photon interactions and the ef\/fective electron mass in a laser background.  We also recall some properties of noncommutative $U(1)$ gauge theories and give our conventions.  In strong f\/ield QED, it is an ef\/fective electron mass which is responsible for much of the new physics which appears relative to QED in vacuum. In Section~\ref{ELECTRON.SECT} we investigate the noncommutative corrections to this ef\/fective mass through the classical Dirac equation and the fermion propagator.  In Section~\ref{PHOTON.SECT} we discuss the propagation of probe photons in a background f\/ield, which is nontrivial since the noncommutative $U(1)$ gauge group is non-Abelian, and make some signif\/icant progress in understanding solutions of the Maxwell equations which include non-constant backgrounds. We will see in our solutions the appearance of both UV/IR mixing and, interestingly, local Lorentz transformations resul\-ting from the presence of star products in gauge transformations. In Section~\ref{SCATTERING.SECT} we examine various strong f\/ield processes in noncommutative QED. We will see that the noncommutative photon plays a dominant role in generating observable corrections to strong f\/ield QED results.

Throughout this paper we use familiar QED f\/ields as our backgrounds, which we often associate with the f\/ields of intense lasers. Since the f\/ield strength $F_{\mu\nu}$ is not gauge invariant in NCQED, though, one might raise the question of whether the physical content of these backgrounds matches what we expect from QED. In Section~\ref{GAUGE.SECT}, therefore, we take some steps toward def\/ining physical, gauge invariant variables of NCQED in conf\/iguration space, and conf\/irm that the physics of our background f\/ields is what one would hope. We conclude in Section~\ref{CONCS.SECT}.

\subsection{Strong f\/ield QED} \label{REVSECT1}
We begin by outlining the Furry picture approach to calculating strong f\/ield QED processes, the ef\/fects of a strong background on the electron, and some experimental signatures. The f\/irst step is to take the QED action and shift the gauge f\/ield as so: $A_\mu\to A_\mu + a_\mu$, so that $A_\mu$ remains the quantum gauge f\/ield and $a$ is a f\/ixed background. The action becomes
\begin{gather}\label{QED.ACT}
	S=\int \ud^4x  -\frac{1}{4} F_{\mu\nu} F^{\mu\nu} + \bar{\psi} (i \slashed{\D}-m)\psi  - e \bar\psi \slashed{A}\psi + \text{gauge f\/ixing} .
\end{gather}
The f\/irst term is the usual kinetic term for the quantum f\/ield\footnote{The kinetic term for the background contributes a constant, which we throw out, and a cross term with $F_{\mu\nu}$, which vanishes after integrating by parts.}
 $A$. The second term contains  $\D_\mu = \partial_\mu + ie a_\mu$, the background covariant derivative, but remains quadratic in $\psi$. These f\/irst two terms are taken to def\/ine the `free' theory, and therefore the propagators. Clearly the photon propagator is unchanged (this will not be the case for NCQED with background f\/ields), while the fermion propagator becomes a `dressed' propagator,
\[
	\frac{1}{i \slashed{\partial}- m + i\epsilon} \to \frac{1}{i \slashed{\D}- m + i\epsilon}.
\]
The third term of (\ref{QED.ACT}) is the normal QED vertex which is treated in perturbation theory as usual. Thus, the Feynman diagrams of the theory have precisely the same form as when the background $a_\mu=0$, but the free (vacuum) fermion propagator is replaced by the dressed propagator which contains all powers of the coupling.  Thus, we calculate in an $\hbar$ expansion rather than a coupling expansion.  Working with such a perturbative series is the functional equivalent of the Furry picture~\cite{Furry:1951zz, MoortgatPick:2009zz}. One reason for adopting the Furry picture is that the background f\/ields we consider here are models of laser f\/ields, which can now reach intensities of around $10^{22}~\text{W}/\text{cm}^2$ and are characterised by parameters (which we will meet shortly) far in excess of unity. Thus, they are not immediately amenable to a perturbative expansion. The Furry picture allows us, in principle, to treat the background exactly. Practically, this is feasible whenever one can solve the Dirac equation in the background f\/ield, essentially, and so construct the dressed propagator.

This brings us to our choice of background f\/ield. We mainly consider `null f\/ields', i.e.\ plane waves $a_\mu\equiv a_\mu(k\sp x)$ such that $k^2=0$ and $k\sp a=0$. In particular, we consider the circularly polarised, linearly polarised and crossed f\/ield potentials
\begin{gather}
	\label{CIRC.POL}	a_\mu(k\sp x)  =  \cos(k\sp x)l_\mu + \sin(k\sp x){\bar l}_\mu  , \\
	\label{LIN.POL}		a_\mu(k\sp x)  =  \cos(k\sp x)l_\mu  , \\
	\label{CROSSED}	a_\mu(k\sp x)  =  k\sp x  l_\mu  ,
\end{gather}
respectively, where $k\sp l= k\sp {\bar l} = 0$. (These may be taken of either f\/inite or inf\/inite duration in $k\sp x$.) The crossed f\/ield potential yields a constant f\/ield strength. Throughout this paper, $k_\mu$ characterises our background f\/ields, and may be thought of as the momentum of photons in a monochromatic laser. The amplitude of our backgrounds is contained in the polarisation vectors~$l_\mu$ and~${l}'_\mu$, which are space-like, of equal magnitude and orthogonal, so $l\sp {\bar l}=0$. The amplitude of the f\/ield can be translated into an intensity parameter called $a_0$, which measures the strength of the background as seen by the electron, and parameterises all intensity ef\/fects~in this paper. For an electron of momentum~$p$, the proper covariant def\/inition is \cite{Heinzl:2008rh}
\begin{gather}
\label{a0.DEF}
	a_0^2 = \frac{e^2}{m^2}\frac{\langle p^\mu T_{\mu\nu} p^\nu \rangle}{(k\sp p)^2},
\end{gather}
where $T_{\mu\nu}$ is the energy momentum tensor and the brackets denote an average over many periods of the laser. Going to the electron rest frame, $a_0$ becomes a ratio of the energy gained by an electron over one period of the laser, to its rest mass. One can easily verify that for the circularly and linearly polarised beams (\ref{CIRC.POL}) and (\ref{LIN.POL}), $a_0 = e |l|/m$ and $a_0 = e |l|/2m$ respectively, essentially the beam amplitude measured in units of~$m$. Physically, f\/ields with an $a_0$ of order $10^2$ are now achievable, while $a_0 = 10^3\sim 10^4$ will be available at the next generation of facilities, such as ELI~\cite{Heinzl:2008an}.

In an oscillating laser f\/ield, an electron undergoes very rapid quiver motion. It is the average momentum over a period, the `quasi' momentum, to which physical processes are most sensitive, as we will shortly see \cite{McDonald}. For an electron of momentum $p_\mu$ incident upon a circularly or linearly polarised laser, the quasi momentum $q_\mu$ is
\begin{gather}
\label{QUASIMOM}
	q_\mu = p_\mu + \frac{a_0^2m^2}{2k\sp p}k_\mu,
\end{gather}
up to a Lorentz transformation depending on the duration of the `pulse' (in $k\sp x$). Squaring the quasi momentum, we f\/ind the well known ef\/fective mass of an electron in a laser background,
\begin{gather}
\label{MASS.SHIFT}
	q^2 = m^2(1+a_0^2) \equiv m_*^2.
\end{gather}
We now investigate some consequences of this mass shift.

\subsubsection*{Nonlinear Compton scattering}

\begin{figure}[t]
\centerline{\includegraphics{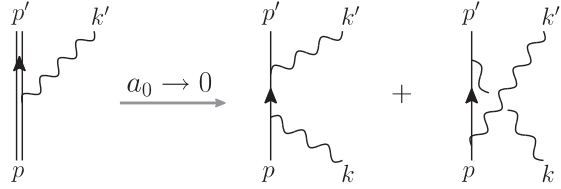}}
\caption{To the left, Furry Feynman diagram for nonlinear Compton scattering of a photon of momentum $k'$, from an electron of momentum $p$. To the right, the low intensity limit reproduces ordinary Compton scattering, multiplied by $a_0^2$ as a f\/lux factor.}
\label{NLC.FIG1}
\end{figure}

Nonlinear Compton scattering is the name given to photon emission by an electron in a laser background (classically, the radiation from an electron accelerated by the background). This is the most experimentally accessible strong f\/ield process, as there is no threshold dependence, so intensity ef\/fects may be observed at whatever $a_0$ is available. There is a single Furry picture diagram contributing to this process at tree level, show to the left of Fig.~\ref{NLC.FIG1}. Double lines indicate dressed propagators throughout this paper. The analogous diagram in vacuum, i.e.\ with undressed fermion lines, vanishes by momentum conservation.
The cross section was f\/irst evaluated for the case of inf\/inite plane waves in \cite{Nikishov:1963, Nikishov:1964a,
 Narozhnyi:1964}. A circularly polarised beam gives the simplest expressions, so we consider that case. Calculating the $S$-matrix element, one f\/inds an inf\/inite sum such that
 \begin{gather}\label{IPW.PROC}
 	\frac{|S_{fi}|^2}{VT} = \sum\limits_{n\geq 1} \delta^4(q_\mu + nk_\mu = q'_\mu + k'_\mu) |M_n|^2.
 \end{gather}
The delta functions conserve the quasi momentum of the colliding particles: $q_\mu$ is the quasi momentum of the incoming electron, see (\ref{QUASIMOM}), and $q'$ that of the outgoing electron (with $p$ replaced by $p'$). Thus, the $S$-matrix element may be interpreted as describing the absorption of $n$ laser photons $\gamma_{\scriptscriptstyle L}$, and the emission of one photon, by a heavy electron of rest mass $m_*$, i.e.\ the processes $e^\m_* + n \gamma_{\scriptscriptstyle L} \to e^\m_* + \gamma$.

\begin{figure}[t]
\centerline{\includegraphics{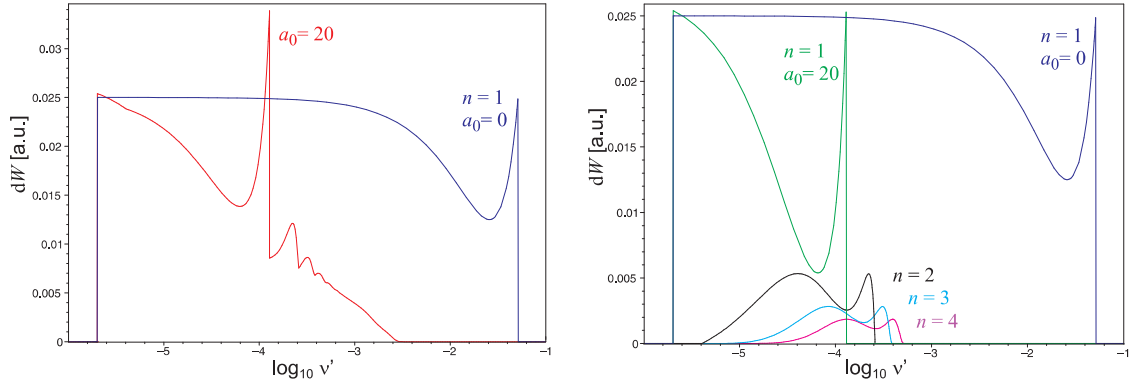}}
\caption{{\it Left:} Dif\/ferential cross section for nonlinear Compton scattering of $80$ MeV electrons from an optical laser of intensity $a_0=20$, plotted as a function of the scattered photon frequency $\nu'\equiv \omega'/m$, and compared to ordinary Compton scattering of the electron with an optical frequency photon at $a_0=0$. {\it Right:} Contributions from individual harmonics which, summed, reproduce the full cross section.}
\label{NLC.FIG}
\end{figure}

It is useful to compare this with ordinary Compton scattering in vacuum, in which an undressed electron scatters of\/f a single photon ($a_0=0$, $n=1$) of momentum $k$, with the scattered photon having momentum $k'$. This is the low intensity limit of our processes, as illustrated in Fig.~\ref{NLC.FIG1}. The ef\/fect of the background f\/ield is twofold. First, the electron recoils less in the collision, due to its increased mass, which decreases the energy transfer to the scattered photon. This results in a redshift of the maximum scattered photon frequency, called the Compton edge, by a factor  $\sim 1/a_0^2$ for high energy incident electrons. This is illustrated in the left panel of Fig.~\ref{NLC.FIG}, with the `nonlinear edge' (in red, to the left) clearly visible and redshifted  relative to the ordinary, or `linear' edge. The increased size of this peak over the linear Compton signal brings us to the second ef\/fect of the background~-- the inf\/inite sum of contributions corresponding to multiple laser photon absorption, or `higher harmonics'.  The f\/irst four contributions are shown in the right panel of Fig.~\ref{NLC.FIG}, with the $n=1$ contribution being dominant, and the higher harmonics $n=2,3,4$ bolstering the signal at the nonlinear edge.  These also contribute the series of smaller maxima corresponding to the scattering of higher energy photons, visible to the right of the nonlinear edge in both panels. These are summed over to arrive at the full rate, in the left panel. Though only a f\/inite number of harmonics are summed in our f\/igure, summing over larger $n$ simply smears the higher harmonic peaks and gives a slower fallof\/f of the emission rate to the right of the f\/igure. Note that the sum over all harmonics converges~\cite{Harvey:2009ry}.

\subsubsection*{Stimulated pair production}
Applying crossing symmetry, we obtain from nonlinear Compton scattering the amplitude for stimulated pair production. Here, a photon of momentum $k'$ is incident upon the laser and produces an electron positron pair. In analogy to the above, the scattering amplitude is an inf\/inite sum over processes $\gamma + n\gamma_{\scriptscriptstyle L} \to e^\m_* + e^\p_*$, i.e.\ a sum over multi-photon Breit Wheeler processes with momentum conservation
\begin{gather}\label{MBW.CONS}
	k'_\mu + nk_\mu = q_\mu + q'_\mu,
\end{gather}
where $q$ and $q'$ are now outgoing quasi momenta for the electron and positron. It follows from squaring (\ref{MBW.CONS}) that the pair production threshold is blue shifted by the ef\/fective mass: the minimum number of photons required from the laser increases according to
\[
	\frac{2m^2}{k\sp k'} \to \frac{2m_*^2}{k\sp k'},
\]
as it is a heavy pair which must be created. It can be shown that in a more realistic background of f\/inite temporal duration, there is a nonzero pair production rate above the ordinary threshold $2m^2/(k\sp k')$, but that this is suppressed up to the ef\/fective mass threshold \cite{Heinzl:2010vg}.

\subsubsection*{Vacuum birefringence}
It is clear from the description of the Furry picture that both real and virtual fermions are dressed by the background f\/ield. The ef\/fect of the latter is observed in loop corrections to the photon propagator:  the dressing of virtual pairs causes the vacuum to develop nontrivial refractive indices, so that the dispersion relation of probe photons depends on their polarisation. This is `vacuum birefringence'  \cite{Toll:1952}, and can be observed in principle by colliding, say,  a linearly polarised beam of photons with a high intensity laser. Emerging from the focal region of the laser, the photon beam will have developed an ellipticity. For more details see \cite{Heinzl:2008an, Heinzl:2006xc}. Vacuum birefringence can also be used as a probe of space-time noncommutativity, see \cite{Abel:2006wj}.

\subsection{Noncommutative QED}\label{NCINTRO.SECT}

\subsubsection*{The action}

We consider noncommutativity generated by the usual Moyal star product,
\[
	f(x) \star g(x) = f(x)\exp{\bigg( \frac{i}{2}{\overset{\leftarrow}{\partial_\mu}}  \theta^{\mu\nu} \overset{\rightarrow}{\partial_\nu}   \bigg)} g(x)  ,
\]
and the NCQED action in vacuum is
\[
	S=\int \ud^4x  -\frac{1}{4} F_{\mu\nu}\star F^{\mu\nu} + \bar{\psi} \star (i \slashed{D}\star-m)\psi+ \text{gauge f\/ixing + ghosts},
\]
where $D_\mu\star$ is the covariant derivative ($D_\mu\star = \partial_\mu + ie A_\mu \star$ in the fundamental) and the f\/ield strength is $F_{\mu\nu} = \partial_\mu A_\nu - \partial_\nu A_\mu + ie \starc{A_\mu}{A_\nu}$. Since we deal almost exclusively with NCQED f\/ields, we drop the `hats' usually used  to separate them from their QED counterparts. When a~QED f\/ield appears, we will be explicit about it.

\subsubsection*{The $\boldsymbol{U(1)}$ gauge group}

Let us recall some basic properties of the f\/ields. The gauge potential and f\/ield strength transform according to
\begin{gather}
\label{A.trans}
	A_\mu^U  \equiv  U^{-1}\star A_\mu \star U+ \frac{1}{ie} U^{-1}\star \partial_\mu U , \\
\label{F.TRANS}
	F_{\mu\nu}^U  \equiv  U^{-1}\star F_{\mu\nu}\star U ,
\end{gather}
where $U$ is a `star-gauge' transformation given by a star exponential,
\begin{gather*}
	U = \exp_\star (ie\alpha) \equiv 1 + ie\alpha +\frac{(ie)^2}{2!}\alpha\star\alpha + \cdots
\end{gather*}
and $U^{-1}= U^\dagger$ is its inverse such that $U^{-1}\star U = 1$.  It is clear from these expressions that NCQED is a non-Abelian gauge theory, with three and four photon vertices appearing in the Feynman rules. Expanding the gauge f\/ields into background and quantum parts will introduce additional vertices, which we will come to later on.

One of the more recent motivations for studying noncommutative f\/ield theory was its appearance as a low energy limit of certain string theories \cite{Seiberg:1999vs}. The hope was that some aspects of the string theory could be understood using the simpler framework of f\/ield theory, and indeed, traits associated with quantum gravity theories certainly persist in noncommutative gauge theory. To see this, consider the gauge transformation $U_c(x) = \exp(i e  c\sp x)$, for a constant vector~$c_\mu$. The nonlocality of the gauge transformations results in a mixing of space-time and internal degrees of freedom, since, using (\ref{A.trans}), the transformed gauge f\/ield is
\begin{gather}
\label{TRANS.GAUGE}
	A^{U_c}_\mu(x) = A_\mu(x - e \theta\sp c) + c_\mu .
\end{gather}
So, up to a global translation, this gauge transformation generates a space-time translation, a property shared with gravitational theories  \cite{Gross:2000ba, Langmann:2001yr}. In Section~\ref{PHOTON.SECT} we will in fact see the appearance of a local Lorentz transformation.

\subsubsection*{Perturbation theory and UV/IR mixing}

There are many aspects of NCQED which we will not be able to cover in great depth here. We refer the reader to \cite{Hayakawa:1999yt, Hayakawa:1999zf} for a thorough introduction to gauge f\/ixing and perturbative calculations, to  \cite{Hewett:2000zp} for collider signatures, and to \cite{Ardalan:2000qk, Nakajima:2001uh} for anomalies. We give here a very brief review of UV/IR mixing as, even though we will only encounter it in a mild form, the topic is central to noncommutative f\/ield theory.

Perturbation theory in NCQED suf\/fers from the UV/IR problem which is ubiquitous in noncommutative f\/ield theories \cite{Minwalla:1999px, Matusis:2000jf}. The free propagators match those of QED, as does overall momentum conservation of Feynman diagrams. The origin of the UV/IR problem is in the interaction vertices, where star products give momentum dependent phase factors between f\/ields in momentum space. The ef\/fect on the Feynman rules is that the momenta f\/lowing into and out of vertices must be ordered, i.e.\ a double line notation is required even for the space-time momenta. This means that diagrams come in two f\/lavours -- planar and nonplanar.  The planar diagrams behave as in QED, in terms of their regularisation and in the limit $\theta\to 0$. The behaviour of the nonplanar diagrams is quite dif\/ferent, however, as they retain a nontrivial phase factor coming from the star products. Introducing a UV cutof\/f $\Lambda$ to regulate the diagrams, one f\/inds that this forces the introduction of an IR cutof\/f which typically goes like $(\theta\Lambda)^{-1}$. Let us brief\/ly illustrate this with the example from \cite{Matusis:2000jf}. The fermion contribution to the one loop photon self energy is
\[
	\Pi_{\mu\nu}(p) = -8 e^2 \int^\Lambda \frac{\ud^4 l}{(2\pi)^4} \bigg(\frac{2l_\mu l_\nu-l^2 g_{\mu\nu}}{l^4}\bigg) +4 e^2 \int^\Lambda \frac{\ud^4 l}{(2\pi)^4} \bigg(\frac{2l_\mu l_\nu-l^2 g_{\mu\nu}}{l^4}\bigg)\exp(i \check{p}\sp l),
\]
where $\check{p}^\mu = \theta^{\mu\nu}p_\nu$ (the gauge boson contribution is similar, see \cite{Matusis:2000jf}). The f\/irst integral above comes from the planar graphs and is the same (up to the prefactor) as that in QED, being quadratically divergent. The second integral is the nonplanar contribution, and because of the oscillating phase factor, is {\it finite}. Removing the cutof\/f, this integral is proportional to
\[
	\frac{\check{p}_\mu \check{p}_\nu}{\check{p}^2},
\]
which diverges quadratically as $\check{p}\to 0$; there is an {\it infrared} divergence. Hence, we do not recover the commutative result as $\theta\to 0$. This is the f\/irst consequence of UV/IR mixing: the limits $\Lambda\to\infty$ (removal of the cutof\/f) and $\theta\to 0$ do not commute.  Worse, when the corresponding diagrams appears in higher loop contributions they lead to divergences for vanishing internal momenta $p\to 0$, and these, being nonlocal, cannot be removed by redef\/initions of the parameters in the action. The theory is nonrenormalisable. (See \cite{Martin:1999aq}, though, for renormalisation to one loop.)  It is natural, then, to look for an extension of the basic Yang--Mills action $\int F\star F$ which will remove the UV/IR mixing.

The UV/IR problem persisted until the authors of \cite{Grosse:2004yu} modif\/ied the standard noncommutative $\phi^4$ theory by including a harmonic term in the kinetic term of the action, replacing $\partial^2$ with $\partial^2 + \omega^2\tilde{x}^2$, where $\tilde{x}_\mu = 2 \theta^{-1}_{\mu\nu}x^\nu$. The ef\/fect of this term is, essentially, to introduce an infrared cutof\/f into the theory which allows a separation of scales, and the theory becomes renormalisable to all orders. (See \cite{deGoursac:2010zb} for a recent discussion of the interpretation of the new term.) Following this advance, external gauge f\/ields were coupled to the renormalisable scalar theories in \cite{de Goursac:2007gq, Grosse:2007dm}. By integrating out the matter f\/ield, one arrives at an ef\/fective action for the external f\/ields. Calculation of the one loop ef\/fective action leads to the following conjecture for a renormalisable gauge f\/ield action:
\[
	\int \ud^4x  -\frac{1}{4} F_{\mu\nu}\star F^{\mu\nu} +\frac{\Omega^2}{4}{\left\{A'_\mu\,\overset{\star}{,}\,A'_\nu\right\}}^2 + \kappa A'_\mu \star {A'}^\mu ,
\]
where $A'_\mu \equiv A_\mu + \tilde{x}_\mu/2$, and $\Omega$, $\kappa$ are parameters. This action is gauge invariant, and the same quadratic terms as in the scalar model of \cite{Grosse:2004yu} appear, which is certainly promising for the purposes of renormalisation. For comprehensive initial investigations of this action see \cite{deGoursac:2009gh}.

Although we will touch upon UV/IR mixing in Sections~\ref{ELECTRON.SECT} and~\ref{PHOTON.SECT}, and it will appear in Section~\ref{SCHWING}, we will not encounter it in very great depth. (It will of course play a central role when the results presented below are extended to higher loops.) Consequently, we will leave our discussion of UV/IR mixing at this point,  referring the reader to \cite{Grosse:2006hh, Rivasseau:2007rz, Blaschke:2009rb, Fischer:2010zg} for more detailed reviews of this important topic and its connection to Langmann--Szabo duality \cite{Langmann:2002cc}, to~\cite{Blaschke:2008yj} for another approach to removing the UV/IR problem, and to \cite{Schupp:2008fs, Raasakka:2010ev} for UV/IR mixing and the Seiberg--Witten map.

\subsubsection*{Notation and conventions}

We end this section with some notation. We use the following wedge product to represent contractions with the noncommutativity tensor,
\[
	k \wp p \equiv \frac{1}{2} k_\mu \theta^{\mu\nu} p_\nu  .
\]
 our background f\/ield is characterised by $k_\mu$, which is light-like, we choose $k_\mu \equiv k_\p \delta_{\mu\p}$, intro\-du\-cing lightcone coordinates
\[
	x^{\LCpm} = x^0 \pm x^3 ,\qquad x^{\LCperp} = \{x^1,x^2\} .
\]
We will see that the noncommutativity tensor always appears in the combination $k_\mu \theta^{\mu\nu} = k_\p\theta^{\p\nu}$, and we assume that $\theta^{\p\m}=0$, in keeping with the discussion in the introduction. Hence, for some momentum $p_\mu$, we have $k\wp p = \tfrac{1}{2} k_\p\theta^{\p\LCperp}p_\LCperp$, which sees the transverse   components of $p_\mu$.

\section{The noncommutative electron}\label{ELECTRON.SECT}

The noncommutative Dirac equation in a background f\/ield $a_\mu$ is
\begin{gather}\label{Dirac}
	i\slashed{\mathscr D}\star\psi - m\psi =0 ,
\end{gather}
where the background covariant derivative is $ \mathscr{D}_\mu\star = \partial_\mu + ie  a_\mu\star$ for fundamental matter and $\mathscr{D}_\mu\star = \partial_\mu + ie  \starc{a_\mu}{-}$ for f\/ields in the adjoint. The case of a plane wave background was studied extensively in \cite{AlvarezGaume:2000bv}. The great simplif\/ication this choice of background af\/fords is that solving~(\ref{Dirac}) becomes equivalent to a well studied commutative problem, as we now describe. The solutions of the commutative Dirac equation in plane waves are Volkov electrons (and positrons) having the form $\exp(-ip\sp x)F(k\sp x)$, at least when one adopts Landau or lightcone gauge~\cite{Volkov:1935}.  Hence, we guess a solution of the same form for our noncommutative problem. The essential result is that the star product between $a_\mu(k\sp x)$ and such a function reduces to an ordinary commutative product,
\[
	a_\mu(k\sp x) \star e^{-ip \cdot x}F(k \sp x) = a_\mu(k\sp x+k\wp p)  e^{-ip \cdot x}F(k\sp x)  .
\]
We therefore need only solve the commutative problem in a background which is shifted by an amount proportional to the electron momentum. If we write $a^\mu_p \equiv a^\mu(k\sp x+k\wp p)$,  then the solution for a fundamental electron, normalised to be free in the inf\/inite past, is
\begin{gather}\label{volky}
	\psi^\m_p(x) = \exp\bigg(-ip\sp x -\frac{i}{2k\sp p}\int^{k\cdot x}_{-\infty} 2e a_p\sp p - e^2a_p^2\bigg) \bigg[ \Eins + \frac{e}{2k\sp p} \slashed{k}\, \slashed{a}_p\bigg] u_p ,
\end{gather}
where it is assumed that $a_\mu\to 0$ as $k\sp x\to-\infty$. The scalar electron is given by dropping all the spinor structure, and the commutative limit $\theta\to 0$ simply replaces $a_p\to a$. For neutral, adjoint matter, on the other hand, the Dirac equation is solved by (\ref{volky}) with $a^\mu_p$ replaced by
\begin{gather}\label{both}
	a^\mu(k\sp x+k\wp p) - a^\mu(k\sp x-k\wp p) .
\end{gather}
These momentum-dependent shifts are interpreted as follows. If the background f\/ield has compact support, then, according to (\ref{volky}), a noncommutative electron sees the pulse a time $k\wp p$ earlier or later than it's commutative counterpart if $k\wp p$ is positive or negative, respectively. The neutral particle, though, sees in (\ref{both}) both an advanced and a retarded f\/ield  and was therefore interpreted in \cite{AlvarezGaume:2000bv} as a dipole, of length $2k\wp p = k \sp \theta \sp p$, with opposite charges at its endpoints. Thus, the adjoint particle behaves as an extended, dipole-like object, a property expected from a stringy perspective where noncommutative f\/ield theories are obtained from decoupling limits, see \cite{Barbon:2001bi}. The charged electron, on the other hand, might be interpreted as a `half dipole' with an ef\/fective length of $k\wp p$. Observe that these solutions exhibit a mixing of UV and IR scales -- as the energy momentum of the electron increases, so does the length of the dipole. This ef\/fect will reappear as UV/IR mixing in the ef\/fective action, which we will encounter in Section~\ref{SCHWING}. From here on we focus on fundamental matter only: see \cite{AlvarezGaume:2000bv} for more details on adjoint matter.

To further investigate this ef\/fective extension, we would like to see it appear in some gauge invariant measure of the background, as seen by the electron. While $\bar\psi\star \psi$ calculated from~(\ref{volky}) matches the QED result, the current $\bar\psi \gamma^\mu \star \psi$ sees the background $a(k\sp x+2k\wp p)$, as is easily checked. Thus, it sees the full length of the electron `half dipole'. We can also construct a measure of f\/ield intensity seen by the electron, which extends the classical intensity parameter $a_0$ in (\ref{a0.DEF}) to NCQED. We expect the intensity of the f\/ield, as seen by the electron, to be proportional to the photon density in the background, which requires an expression quadratic in $F_{\mu\nu}$. We then have to mop up two vector indices and the star-gauge transformation of, essentially, $F\star F$.  This naturally points us at the energy momentum tensor \cite{Gerhold:2000ik, Das:2002jd} as in (\ref{a0.DEF}), but this reduces to $F\star F$ in our plane waves. We are lead to def\/ining a parameter, $\eta$, as so,
\[
	\eta^2(k\sp x) := \frac{e^2}{m^2}\frac{   \overline{D^\mu\psi} \star F_{\mu\sigma}\star {F^\sigma}_\nu \star D^\nu \psi  }{(k\sp p)^2} ,
\]
which is made dimensionless as in (\ref{a0.DEF}). This is a Lorentz covariant, gauge invariant and local object\footnote{Recall that in noncommutative gauge theories no local gauge invariant observables can be constructed from the gauge f\/ield alone, in coordinate space, but here we also use the matter f\/ields. We will return to this point in Section~\ref{GAUGE.SECT}.}. For a plane wave with potential $a_\mu(k\sp x)$, we f\/ind (with the spinors normalised to unity)
\begin{gather*}
	 \eta^2(k\sp x) = \frac{e^2 {a'}^2(k\sp x+2 k\wp p)}{m^2} .
\end{gather*}
Again, because the expression is quadratic in $\psi$, the shift in the argument of the background is twice that appearing in the Volkov electron, i.e., this gauge invariant measure also sees a~displacement we can associate with the full length of the electron `half dipole'. Our $\eta$ is a~scalar f\/ield, which, since it is already gauge invariant, we can make into a scalar value by, say, ave\-ra\-ging over many wavelengths of the laser. In this case, it is clear that when $\theta\to 0$, we will have $\langle\eta\rangle\to a_0$, the commutative intensity parameter. Indeed, if we consider a periodic plane wave, then we f\/ind $\langle \eta\rangle=a_0$, so that the f\/ield intensity seen by the electron here is the same as in the commutative case.

\subsection{The ef\/fective electron mass}\label{PROP.SECT.1}

The electron propagator is easily obtained from the Volkov solutions using the spectral representation. It is suf\/f\/icient for our purposes here to consider only the scalar  propagator, which also keeps things simple. We f\/ind, either from the spectral sum or direct calculation \cite{Brown:1964zz, Reiss:Greens, AlvarezGaume:2000bv},
\begin{gather}\label{SCALAR.PROP}
	G(x,y) = \int \frac{\ud^4 p}{(2\pi)^4}  \frac{e^{-ip\cdot (x-y)}}{p^2-m^2+i\epsilon}\ \exp\bigg[-\frac{i}{2k\sp p}\int^{k\cdot x}_{k\cdot y} 2e a_p\sp p - e^2a_p^2 \bigg] .
\end{gather}
This clearly reduces to the free propagator when the background is switched of\/f. When the f\/ield is present, though, the structure of the propagator is very dif\/ferent\footnote{This propagator may also be constructed using the f\/irst quantised noncommutative theory of \cite{Dymarsky:2001xg}, which extends the constant background calculations of that paper.}. To illustrate, consider the circular and linear potentials (\ref{CIRC.POL}) and (\ref{LIN.POL}). Then the integrand in the exponent of (\ref{SCALAR.PROP}) may be expanded into an oscillatory piece and a constant piece. Expanding all the oscillatory dependence of the propagator in a double Fourier series allows us to write
\[
	G(x,y) = \int \frac{\ud^4 p}{(2\pi)^4}   \frac{e^{-iq\cdot (x-y)}}{p^2-m^2+i\epsilon}  \sum\limits_{n,n'} \Gamma_n\bar{\Gamma}_{n'} e^{-ink \cdot x+in'k\cdot y} ,
\]
where $q_\mu$ is the quasi momentum from (\ref{QUASIMOM}). This receives no noncommutative corrections, which appear instead in the Fourier coef\/f\/icients $\Gamma$. We will not need their explicit forms here; the double series is necessary because the background breaks translation invariance (in the $k\sp x\sim x^\p$ direction). Fourier transforming to momentum space proper, $x\to p$, $y\to p'$, we see that a~typical term of the sum has the important structure
\[
	\widetilde{G}(p,p') \sim \delta^4(p-p' + n k - n' k)  \frac{1}{(p+nk)^2-m_*^2+i\epsilon} ,
\]
where $m_*$ is just as in (\ref{MASS.SHIFT}). The simplest term has $n=n'=0$, for which we have ordinary, momentum conserving propagation, but for a particle of mass $m_*$. The remaining poles describe propagation during which  the particle emits or absorbs real photons of momentum $k_\mu$ (which is possible because the electron is immersed in a background, not in vacuum).   Importantly, the ef\/fective electron mass is unchanged by noncommutative ef\/fects. This is because, in a periodic beam, the shift $a^\mu\to a^\mu_p$ does not af\/fect the average momentum of a particle, so neither the quasi momentum nor the mass shift receive noncommutative corrections. Even though this will not hold in more general backgrounds, we note that for f\/inite duration plane wave backgrounds which are periodic, or nearly so, scattering processes exhibit a resonant behaviour at energy transfer corresponding to the interactions of ef\/fective mass electrons \cite{Heinzl:2010vg}, so the dominant contributions to dif\/ferential rates come from inf\/inite plane wave processes, e.g.~(\ref{IPW.PROC}). Hence, dealing with these simple backgrounds is a good f\/irst approximation. We will see some consequences of the unchanged mass shift in Section~\ref{SCATTERING.SECT}.

\section{The noncommutative photon} \label{PHOTON.SECT}

The noncommutative Maxwell equations are
\begin{gather} \label{full-Max}
	D^\mu\star F_{\mu\nu} = 0 ,
\end{gather}
where the covariant derivative is $D_\mu\star \equiv \partial_\mu + ie \starc{A_\mu}{-}$. A transverse plane wave remains a~solution of (\ref{full-Max}) in NCQED (compare the plane wave solutions in ordinary Yang--Mills theory~\cite{Coleman:1977ps}), as does a constant f\/ield strength. In the latter case, the f\/ield strength is gauge invariant, as is apparent from (\ref{F.TRANS}). We now want to consider superpositions of backgrounds~$a_\mu$ and probes~$B_\mu$, so that the background is decomposed as $A\equiv a+B$. We begin by reviewing the known solutions in a constant background. We then give a revealing new parameterisation of these solutions and extend this to the case of of nonconstant backgrounds.

\subsection{Constant backgrounds}
In \cite{Mariz:2006kp} it was shown that a superposition of a constant f\/ield strength and an arbitrary plane wave is an exact solution to (\ref{full-Max}). To illustrate, take the crossed f\/ield potential (\ref{CROSSED}), giving us a constant f\/ield strength (see \cite{Guralnik:2001ax} for the case of a constant magnetic background), and $B_\mu\equiv B_\mu(p\sp x)$ a~plane wave depending on some vector $p_\mu$. Since this is univariate, all star-commutators of~$B$ with itself vanish. Consequently, although the plane wave will interact with the background f\/ield, there will be no self interaction, and~$B_\mu$ need only satisfy the linearised version of (\ref{full-Max}), which imposes two constraints:
\begin{gather}\label{old}
	\bar{p}\sp \bar{p} = 0 ,\qquad \bar{p} \sp B = 0 ,
\end{gather}
where $ \bar{p}_\mu \equiv p_\mu - e l_\mu  k\sp\theta\sp p$.   These equations strongly resemble those on transverse plane waves, but they are phrased in terms of a shifted momentum $\bar{p}_\mu$ which sees the noncommutativity. The simplicity of these constraints is attractive, but the solutions of the dispersion relation $\bar{p}^2=0$ are not transparent, nor does (\ref{old}) give us an immediate description of how the polarisation vector is af\/fected \cite{Chatillon:2006rn}. So, let us instead give a parameterisation of the solution in terms of wave- and polarisation vectors for commutative f\/ields. This will not only shed light on the meaning of the solution but, as we will see, will also be immediately extendable to solutions of (\ref{full-Max}) which include non-constant backgrounds. We f\/irst observe that, expanding $A=a + B$ in (\ref{full-Max}), the constraints in (\ref{old}) come from the two conditions
\begin{gather} \label{21}
	\D^2 B _\mu = 0 , \qquad \D\sp B = 0 ,
\end{gather}
where $\mathscr{D}_\mu$ is the adjoint background derivative introduced previously.  Now, let $k'_\mu$ and  $\epsilon_\mu$ be wave and polarisation vectors for a transverse plane wave, i.e.\ $k'\sp k' = k'\sp\epsilon=0$. These give a~basis $\exp(-ik'\sp x)\epsilon_\mu$ of solutions to the Maxwell equations in QED and in NCQED when the background is switched of\/f. When it is turned on, these basis elements becomes
\begin{gather} \label{sol1}
	\exp\bigg( -ik'\sp x - \frac{i}{2k\sp k'}(2e \delta a\sp k' - e^2 \delta a^2)k\sp x\bigg)  \exp\bigg[\frac{e}{k\sp k'}k^{[\mu}\delta a^{\nu]}\bigg]\epsilon_\nu ,
\end{gather}
where the background seen by the plane wave is $\delta a_\mu$ def\/ined by
\begin{gather} \label{da}
	\delta a_\mu(k\sp x) := a_\mu(k\sp x + k\wp k') - a_\mu(k\sp x - k\wp k')  .
\end{gather}
This is naturally the same background seen by adjoint matter in (\ref{both}). Note that we will use this def\/inition of $\delta a_\mu$ for  arbitrary plane wave backgrounds below, but for crossed f\/ields it reduces to a constant,
\[
	\delta a_\mu\to 2 k\wp k'  l_\mu .
\]
The noncommutative photon, therefore, exhibits the same dipole-like structure described above.   The tensor in (\ref{sol1}) is the exponential of an antisymmetric tensor and is in fact a {\it Lorentz transformation} acting on the polarisation vector, giving a longitudinal (i.e.~$k_\mu$ direction) boost and a transverse rotation. To see this, one can expand the exponential, which terminates at second order, f\/inding
\[
	\exp\bigg[\frac{e}{k\sp k'}k^{[\mu}\delta a^{\nu]}\bigg] = \eta^{\mu\nu} + \frac{e}{k\sp k'}(k^\mu \delta a^\nu - \delta a^\mu k^\nu) - \frac{1}{2}\frac{e^2\delta a^2}{(k\sp k')^2}k^\mu k^\nu   .
\]
This expression, and (\ref{sol1}), are highly reminiscent of the Volkov solution encountered earlier. Just as electrons are `dressed' by the background f\/ield, a gauge boson probe introduced into the background also becomes dressed, as the $U(1)$ star-gauge group is non-Abelian. This generates a shift in the wave vector and a boost of the polarisation, as photons from the background are dragged along by the probe.  Let us investigate this dressing further for nonconstant backgrounds.

\subsection{Non-constant backgrounds: a Furry picture}

We now want to consider what happens when the background f\/ield is a nonconstant plane wave with potential $a_\mu\equiv a_\mu(k\sp x)$ into which we add a probe $B_\mu(x)$. Expanding $A=a + B$ again, the Maxwell equations (\ref{full-Max}) for $B_\mu$ become
\[
	\mathscr{D}^2 B_\mu - \mathscr{D}^\nu  \mathscr{D}_\mu  B_\nu +  i e  \starc{{f_\mu}^\nu}{B_\nu} =
	 i e  \mathscr{D}^\nu  \starc{B_\mu}{B_\nu} + i e \starc{\D_{[\nu} B_{\mu]}}{B^\nu}  + e^2 \starc{B_\nu}{\starc{B^\nu}{B_\mu}},
\]
where $f_{\mu\nu}$ is the f\/ield strength calculated from $a_\mu$. (Dropping the right hand side and the commutator with $f_{\mu\nu}$, we f\/ind the constraints (\ref{21}) of the crossed f\/ield case.) The left hand side of this equation is linear in $B$, while the right hand side contains all the nonlinearities. One could solve this equation in a coupling expansion, or in powers of the noncommutativity tensor. More is gained, though, by mimicking the Furry picture approach of the quantum theory. To do so, we {\it forget} about all factors of $e$ which appear only because of the background f\/ield (imagine absorbing it into $a_\mu$ for this discussion). So, the only factors of $e$ we care about appear through the nonlinearities of the Maxwell equations (the self interaction of the photon). We now expand~$B$ in powers of the coupling,
\[
	B_\mu = \sum\limits_{n=0} e^n B^n_\mu ,
\]
insert this into the Maxwell equations, and solve order by order. To illustrate, the f\/irst order equation is
\begin{gather}\label{first}
	\mathscr{D}^2 B^0_\mu - \mathscr{D}^\nu\mathscr{D}_\mu B^0_\nu +ie \starc{{f_\mu}^\nu}{B^0_\nu} = 0,
\end{gather}
while higher order equations take the form
\begin{gather}\label{higher}
	\mathscr{D}^2 B^n_\mu - \mathscr{D}^\nu\mathscr{D}_\mu B^n_\nu +ie\starc{{f_\mu}^\nu}{B^n_\nu}= S^n_\mu,
\end{gather}
where $S^n$ is a source term constructed from the $B^r_\mu$ with $r<n$. In this way, we treat the self-interaction of the probe (the nonlinearity) as a perturbation of the system (\ref{first}), which describes the interaction of the probe and the background. This linearised equation (\ref{first}) can again be solved exactly in terms of commutative variables $k'$ and $\epsilon$, after which (\ref{higher}) can be solved recursively.  This approach gives the classical analogue of calculating propagators exactly in the background f\/ield, and treating all other interactions in perturbation theory, i.e.\ using the Furry picture. We guess a solution of the form $\exp(-ik'\sp x)F(k\sp x)$, and observe that
\begin{gather*}
	\D_\mu \star e^{-ik'\cdot x} F(k\sp x)  = (\partial_\mu + ie\delta a_\mu)  e^{-ik'\cdot x} F(k\sp x),
\end{gather*}
where $\delta a(k\sp x)$ is def\/ined in (\ref{da}). So again, all star products reduce to commutative products. The general solution to the linearised equations is then most easily found by adopting background covariant gauge $\D\sp A=0$ \cite{Abbott:1981ke}. Since $\D\sp a=0$, this implies $\D\sp B=0$, and we f\/ind
\begin{gather}
	B^{0\mu} = \int  \frac{\ud^3k'}{(2\pi)^3}  \exp\bigg(  -ik'\sp x - \frac{i}{2k\sp k'}  \int^{k\cdot x} 2e \delta a\sp k' - e^2 \delta a^2\bigg) \nonumber\\
  \phantom{B^{0\mu} =}{}\times \exp  \bigg[  \frac{e}{k\sp k'} k^{[\mu} \delta a^{\nu]}  \bigg]  \epsilon_\nu(k') + \text{c.c.},\label{sol2}
\end{gather}
where a sum over polarisations is implied and recall that $\delta a$ is no longer constant. Thus we see one attractive feature of (\ref{sol1})~-- it generalises naturally  to the case of a nonconstant background f\/ield. The links with the Volkov solution (\ref{volky}) are manifest here.  Our  probe photon of momentum~$k'$ is, neglecting the self interaction, dressed by the background f\/ield, acquiring a momentum and polarisation shift. It sees the background (\ref{da}) and behaves as a dipole of length $2k\wp k'$.    Intriguingly, and in contrast to the case of a constant f\/ield strength, the tensor exponential, which again terminates at second order, generates here a {\it local}, i.e.\ gauged, Lorentz transformation, consisting of a longitudinal boost and transverse rotation. This is rather interesting in the light of~(\ref{TRANS.GAUGE}), and we will see some physical ef\/fects of the polarisation transformation in Section~\ref{SCATTERING.SECT}.

\subsubsection*{Nonlinear terms}

Let us now brief\/ly consider the form of the higher order terms which result from the self-interaction of the probe. To do so, we will drop the integral in (\ref{sol2}), focussing on a single mode of the f\/ield. Let us also compactify notation somewhat, writing the phase and the tensor terms together as $M(k',k\sp x)_{\mu\nu}$, so that our f\/irst order solution (\ref{sol2}) becomes
\[
	\tilde{B}^0_\mu = e^{-ik'\cdot x} M(k',k\sp x)_{\mu\nu}\epsilon^\nu + \text{c.c.}
\]
To f\/ind the $n^\text{th}$ order solution $B_\mu^n$, we note that the source terms $S^n$ are constructed from the~$B^r$ with $r<n$  and take the form
\[
	S^n\sim e^{\pm i r k'\cdot x}  S_{n,r}(k\sp x),
\]
with $r \leq n+1$. Using this, the dif\/ferential equation (\ref{higher}) takes a standard form, and its solutions are
\[
	\tilde{B}^{n,r}_\mu = \frac{i}{2 r k' \sp k }  e^{-i r k'\cdot x}{M(r k',k\sp x)_\mu}^\sigma \int^{k\cdot x} \ud y\, M^{-1}(rk',y)_{\sigma\rho} S^\rho_{n,r}(y) + \text{c.c.}
\]
Thus, the self interaction of the probe generates terms of a similar form to (\ref{sol2}), with the same phase and polarisation structures appearing through $M$. Note that the term carrying momentum~$r k'_\mu$, in the leading exponent, sees advanced and retarded backgrounds
\[
	a_\mu(k\sp x + r k\wp k') - a_\mu(k\sp x - r k\wp k').
\]
We can therefore interpret the nonlinearity of our equations as introducing probes of momenta~$rk'_\mu$, which behave as dipoles of length $2 r k\wedge k'$. In nonconstant backgrounds, then, a~probe photon is required to propagate as part of a collection of dipole like objects carrying multiples of a fundamental momentum~$k'_\mu$. As the energy $r |\mathbf{k}'|$ increases, so does the length $2r k\wedge k'$ of the dipole, so that our solution to the Maxwell equations exhibits the UV/IR mixing typical to noncommutative theories.  Given that the gauge group is non-Abelian, further investigation of these, and similar solutions may of\/fer progress into how bound states, or `photoballs' are formed in NCQED \cite{Fatollahi:2005ri, Volkholz:2005mp}. Although we will touch on connected ideas in Section~\ref{GAUGE.SECT}, we leave further study of the nonlinear terms here. The reason is that, in the sequel, we will require only the photon propagator in a background f\/ield, which is constructed from the linearised solution~(\ref{sol2}), and to which we now return.

\subsection{The photon dispersion relation}

The photon dispersion relation in NCQED potentially suf\/fers from instabilities which have two origins. The f\/irst is UV/IR mixing stemming from loop ef\/fects \cite{Hayakawa:1999yt, Hayakawa:1999zf, Matusis:2000jf, Gomis:2000pf} and the second is unitarity problems caused by time-like noncommutativity \cite{Mariz:2006kp} (or, more particularly, improper quantisation of theories with time-like noncommutativity \cite{Bahns:2002vm}). To close this section, we conf\/irm the lack of the latter problem in the case of light-like noncommutativity and nonconstant backgrounds.

Taking our general linear solution (\ref{sol2}), we may calculate the (Furry picture proper) propagator of our dressed photons from a spectral sum. As in Section~\ref{PROP.SECT.1}, we can consider a periodic background to simplify the analysis. Doing so, the propagator will again have an oscillatory structure up to an exponential term which shifts the momentum $k'$ to%
\begin{gather}\label{PHOTON.QUASI}
	k'_\mu \to k'_\mu + \frac{2a_0^2m^2\sin^2(k\wp k')}{k\sp k'}k_\mu .
\end{gather}
This is the photonic analogue of the electron quasi momentum, as is conf\/irmed by Fourier transforming the propagator, $x\to k'$, $y\to k''$. Omitting the details, we f\/ind a sum over terms with the pole structure
\[
	\delta^4(k'-k'' + nk - n'k) \frac{1}{(k'+nk)^2 - 4m^2\Delta^2+i\epsilon},
\]
where we have introduced
\begin{gather*}
	\Delta \equiv \Delta(k') = a_0 \sin(k\wp k') .
\end{gather*}
Consider again the term with $n=n'=0$, then the dispersion relation reads:
\[
	{k'}^2 = 4m^2\Delta^2 = 4a_0^2m^2\sin^2(k\wp k'),
\]
as would also follow from (\ref{PHOTON.QUASI}), so that $2m\Delta$ plays the role of a f\/ield and momentum dependent self energy, or `photon mass'.   Note that $a_0m\sim e|l|$ is actually independent of $m$, so that this momentum has nothing to do with the electron~-- it is simply convenient to continue using $a_0$ to label intensity ef\/fects. Writing $k'_\mu\equiv (\omega',\mathbf{k}')$ the dispersion relation above implies a positive energy
\[
	 E'  = \sqrt{|\mathbf{k}'|^2 + 4a_0^2m^2\sin^2(k_+\theta^{+\perp}\mathbf{k}'_\perp/2)},
\]
assuming $\theta^{+-}=0$ for light-like noncommutativity, so that we can indeed think of $2m\Delta$ as a~photon mass dependent on its transverse momenta. There are again an inf\/inite number of further poles in the propagator, corresponding to multiple emission/absorption of momentum $k_\mu$ from the background through the three and four point vertices. We will conf\/irm later that these are correctly contained in our noncommutative photon.  Now that we have some understanding of the dressed electron and photon, we proceed to discuss various scattering processes in which they appear.

\section{Strong f\/ield,  noncommutative processes} \label{SCATTERING.SECT}

Scattering processes in NCQED will naturally be expected to dif\/fer from QED results both because of noncommutative corrections in the three point vertex, $\bar\psi\star\slashed{A}\star\psi$, and because of the presence of additional scattering channels involving the three-photon and four-photon vertices.  Despite this, strong f\/ield processes often show a lack of decoupling from ordinary QED results. We will give an explanation for this as we progress, and we will see that a key role in generating noncommutative corrections is played by the presence of our noncommutative photon. As a~useful reference, we begin by very brief\/ly reviewing Breit--Wheeler pair production and Compton scattering in vacuum, before reintroducing our background f\/ield.

\subsection[Breit-Wheeler pair production]{Breit--Wheeler pair production}

\begin{figure}[t]
\centerline{\includegraphics{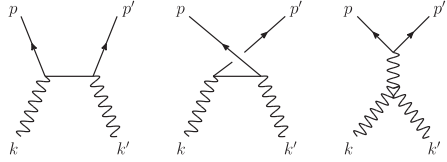}}
\caption{Breit--Wheeler pair production in NCQED.}
\label{BW.FIG}
\end{figure}

The production of an electron positron pair, momenta $p$ and $p'$, through the two photon Breit--Wheeler process is described by the three diagrams\footnote{It is customary to use double lines for noncommutative propagators, as it is necessary to order the momenta f\/lowing in and out of vertices. This is discussed thoroughly in the literature, so we reserve the double line notation for dressed propagators.} of Fig.~\ref{BW.FIG}, which includes an additional scattering channel involving the three-photon vertex.  The usual Mandelstam variables are $\hat{s}=(k+k')^2$, $\hat{t}=(p-k)^2$ and $\hat{u}=(p'-k)^2$. Assuming the collision energy to be well above the pair creation threshold (typically, TeV scales are envisioned) we may neglect the fermion mass, and then the dif\/ferential cross section is proportional to \cite{Baek:2001ty, Godfrey:2001yy, Godfrey:2001sc, Ohl:2004tn}
\begin{gather}\label{pp_amp}
	\ud\sigma \sim \frac{1}{2}\left(  \frac{\hat{u}}{\hat{t}} +  \frac{\hat{t}}{\hat{u}}  -4  \frac{\hat{t}^2+\hat{u}^2}{\hat{s}^2} \sin^2(k\wp k')  \right) =  (2u-1)\bigg( 1 - \frac{\sin^2(k\wp k')}{u}  \bigg)   .
\end{gather}
In the second equality we have introduced the invariant
\begin{gather}\label{u-def}
	u=\frac{(k\sp k')^2}{4k\sp p k\sp p'},
\end{gather}
which will appear again. Noncommutative ef\/fects appear only in the $\sin^2$ term, which has an overall minus sign and thus the cross section is reduced, relative to QED, through destructive interference between the three possible channels.

\subsection{Compton scattering}

The cross section for Compton scattering is obtained through crossing symmetry of the pair production amplitude: send $k'\to -k'$ for an outgoing photon, $p'\to -p$ for an incoming electron and relabel $p\to p'$ for the outgoing electron. One obtains
\[
	\ud\sigma \sim \frac{1}{2} \left( -\frac{\hat{u}}{\hat{s}} - \frac{\hat{s}}{\hat{u}} + 4\frac{\hat{s}^2+\hat{u}^2}{\hat{t}^2} \sin^2(k\wp k')\right) ,
\]
where the Mandelstam variables are now $\hat s = (p+k)^2$, $\hat t = (p'-p)^2$ and $\hat u = (k'-p)^2$. Thus, noncommutative ef\/fects increase the cross section in this case. In these expressions, $k'$ is the momentum of the produced photon, i.e.\ an outgoing momentum variable. The dependence on $k\wp k'$ therefore leads to new angular dependencies in the dif\/ferential rates, relative to QED. This gives us a signal with which to probe noncommutativity and generate bounds on the components of $\theta^{\mu\nu}$. As reported in~\cite{Godfrey:2001yy}, exploiting these angular dependences generates sronger bounds on the noncommutativity scale than considering the totally integrated cross section alone.

Observe that in both Compton scattering and pair production, the noncommutative corrections depend only on the momentum of the photon, but not that of the fermions.

\subsection{Schwinger pair production} \label{SCHWING}

\begin{figure}[t]
\centerline{\includegraphics{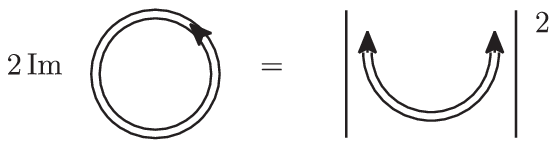}}
\caption{Schwinger pair production, via the optical theorem.}
\label{VACBUB.FIG}
\end{figure}

We now reintroduce our background f\/ield. Recall the discussion of how one calculates amplitudes in a background, as given in Section~\ref{REVSECT1}. Separating the gauge f\/ields into a background and quantum f\/luctuation has two ef\/fects on the NCQED action. First, one obtains corrections to the quadratic terms which dress the propagators. Second, one obtains new vertices between the background and the quantised f\/ields, due to the terms cubic and quartic in the gauge potential. These, along with the original vertices of the theory, are treated in perturbation theory as usual. For the processes we consider here, the only contributing vertices are those of ordinary NCQED, though, so one need only remember to replace propagators by their dressed counterparts. We continue to illustrate this in diagrams by drawing double lines.

We now proceed to our f\/irst process in a background f\/ield -- Schwinger pair production, beginning with the case of a plane wave background. In QED, there is no Schwinger pair production in plane waves, essentially because the magnetic f\/ield works against the electric f\/ield and completely cancels its ``pair creativity''. This result is also characterised in a Lorentz invariant way by the vanishing of both f\/ield invariants,
\[
	-\frac{1}{4}F_{\mu\nu}F^{\mu\nu}=\E^2-\B^2,\qquad -\frac{1}{4}F_{\mu\nu}\tilde{F}^{\mu\nu}=\E\sp \B.
\]

\subsubsection*{The ef\/fective action and UV/IR mixing}
The calculation of the pair creation rate was extended to NCQED in \cite{AlvarezGaume:2000bv}, by calculating the one loop ef\/fective action, i.e.\ the vacuum bubble on the left hand side of Fig.~\ref{VACBUB.FIG}. A noticeable feature of this calculation is that UV/IR mixing appears as a consequence of the electron's extension, as we will now see. Using the worldline formalism, the ef\/fective action reduces to
\[
	\int \ud^4 \, x \delta\mathcal{L}(x) = \int_0^\infty \frac{\ud s}{s^2}\,  e^{-sm^2} \int \ud^4x \int \frac{\ud^2 p_\perp}{(2\pi)^2}  \exp\bigg[-s\big(p_\perp+e a_\perp(k\sp x + k\wp p)\big)^2\bigg] -(a = 0) .
\]
In order to regulate the co-ordinate integral, an infrared cutof\/f is imposed on space-time. Suppose we restrict $-L/2 \leq x^+ \leq L/2$. Then, because of the shift $k\sp x\to k\sp x + k\wp p$ in the argument of the background f\/ield, we are forced to introduce a cutof\/f on the momentum space integration (so that the background waves cannot propagate outside the allowed space-time volume). The restriction on the momentum integral is
\[
	-\omega L \leq k\wp p \leq \omega L,
\]
which is an {\it ultraviolet} cutof\/f on $p$, inversely proportional to $\theta$, as expected from the general discussion of Section \ref{NCINTRO.SECT}. This leads to a quadratic divergence of the form $L^2/|\theta|^2$ in the large volume limit. Such ef\/fects will clearly be generally present in loop diagrams. As they originate from the ef\/fect of the background f\/ield, they are in addition to the usual UV/IR divergences of perturbation theory. It would be extremely interesting to investigate such background-induced ef\/fects in the context of the proposed renormalisable models of \cite{de Goursac:2007gq, Grosse:2007dm}, especially since these were developed through the inclusion of background gauge f\/ields into matter theories. Returning to our ef\/fective action, it ultimately reduces to a boundary term, and there is no Schwinger production in a plane wave, just as in QED. We do not reproduce the whole calculation here, but give instead an alternative derivation of this result from a tree level perspective.

\subsubsection*{The cut diagram and momentum conservation}

One can calculate the Schwinger pair production amplitude directly by evaluating the amplitude on the right hand side of Fig.~\ref{VACBUB.FIG}.  The tree level $S$-matrix element $S_{spp}$ is a double amputation of the dressed fermion propagator, and therefore takes the form
\[
	S_{spp}=\int \ud^4x  \, e^{i{p'}\cdot x+ip\cdot x} F_{pp'}(k\sp x),
\]
where $F_{pp'}$ is a combination of the Volkov spinors for the produced pair of momenta $p$ and $p'$. Its form is not important, as it is straightforward to see from this expression why there is no pair production: taking a Fourier transform with respect to $k\sp x$, we f\/ind
\[
	S_{spp}
	= (2\pi)^3\int \ud s\,  \delta^4(p'+p-sk)\tilde{F}_{pp'}(s) = 0.
\]
This vanishes because, as usual, the NCQED diagram inherits its momentum conservation law from QED, and in this case the delta function has no support. There is no Schwinger pair production in a plane wave, conf\/irming the result of \cite{AlvarezGaume:2000bv}.  In the long wavelength limit, this discussion includes the case of constant f\/ield strength, i.e.\ crossed f\/ields. (A constant electric f\/ield, on the other hand, would be characterised by a time-like vector instead of the light-like~$k_\mu$, which would allow momentum conservation to be preserved. We return to this below.)

Recall that for plane wave backgrounds, the ef\/fective fermion mass receives no contribution from noncommutativity. Hence, in scattering, especially when we average out spin ef\/fects, it seems that we should not expect large deviations from strong f\/ield QED to come from the fermions. This certainly seems to be the case in Schwinger pair production in plane waves, and we will see a second example in a moment. Photons, though, change signif\/icantly in character due to combined noncommutative and background ef\/fects, and we might hope that some trace of this  remains in our cross sections. The ef\/fect of a virtual photon can be explored using constant, but not crossed, f\/ields, to which we now turn.

\subsubsection*{Constant f\/ields}

It has long been known that the vacuum is unstable to pair production in the presence of strong electric f\/ields \cite{Sauter:1931zz, Schwinger:1951nm}, and this remains true in NCQED.  The authors of \cite{Chair:2000vb} considered the noncommutative pair production rate in a constant background f\/ield with nonvanishing invariant $-\tfrac{1}{4}F_{\mu\nu}\tilde{F}^{\mu\nu}=\E\sp \B$, which distinguishes this case from that of crossed f\/ields, included in the plane wave analysis above. For such a background, there exists a frame in which the electric and magnetic f\/ields are parallel. (This is thus more general than Schwinger's original calculation, which is recovered upon taking the magnetic f\/ield strength to zero.) One f\/inds that the pair production rate in such a background is identical to the QED result at tree level, i.e.
\begin{gather}\label{sch}
	\sigma = \frac{\alpha EB}{\pi}  \coth{\bigg(\frac{\pi B}{E}\bigg)}\exp\bigg[-\frac{\pi m^2}{eE}\bigg],
\end{gather}
where $E$ and $B$ are the background electric and magnetic f\/ield strengths. In contrast to the case of crossed f\/ields, a magnetic f\/ield parallel to the electric increases the size of the cross section. There is no change to the Schwinger--Sauter threshold, and the rate remains exponentially suppressed. The same result is found using the f\/irst quantised approach of~\cite{Dymarsky:2001xg}. Adding loop corrections, though, one f\/inds that the exponential suppression factor in~(\ref{sch}) receives a~noncommutative correction \cite{Chair:2000vb,Riad:2000vy},
\[
	\exp\bigg[-\frac{\pi m^2}{eE}\bigg]\to \exp\bigg[-\frac{\pi m^2}{eE}\bigg(1-\frac{e\alpha\gamma }{3\pi}{\bm\theta}\sp {\bm B}\bigg)\bigg] ,
\]
where $\gamma$ is the Euler constant. Thus, the pair production threshold is lowered by the contribution of a virtual, suitably dressed, photon at one loop, and we have our f\/irst strong f\/ield noncommutative correction. We now go on to consider processes with a real photon.

\subsection{Stimulated pair production}

\begin{figure}[t]
\centerline{\includegraphics{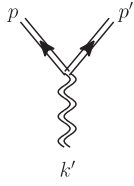}}
\caption{Stimulated NCQED pair production in a background f\/ield.}
\label{NC.3PT}
\end{figure}

There is no Schwinger pair production in (near) null f\/ields, even at high energies. In order to assist, or stimulate, pair production, a high energy probe photon is introduced into the background, and pairs are created through the channel shown in Fig.~\ref{NC.3PT}.  The equivalent diagram in vacuum, i.e.\ with undressed propagators, vanishes in both QED and NCQED by momentum conservation, in just the same way as the Schwinger amplitude vanishes in a plane wave. As with our classical calculations, it is easiest to work in background covariant gauge, in which the necessary amputated propagators in Fig.~\ref{NC.3PT} are given in Sections~\ref{ELECTRON.SECT} and~\ref{PHOTON.SECT}.

The $S$-matrix element is
\[
	S_{fi} = -ie\int \ud^4x \,  \bar{\psi}^{\m}_p(x)\star \slashed{\tilde B}_{k'}(x)\star \psi^{\p}_{p'}(x) ,
\]
where ${\tilde B}_\mu$ is a Fourier mode of (\ref{sol2}), i.e.\ a noncommutative photon wavefunction, and $\psi^{\LCpm}$ are noncommutative positron and electron wavefunctions, respectively.  The star products above are easily evaluated for arbitrary plane wave backgrounds since each function is univariate in $k\sp x$, up to exponential, translating terms. The resulting expressions are somewhat complicated in general, so we give here the simplest case, that of a circularly polarised plane wave of inf\/inite extent.  One f\/inds in this case  that the $S$-matrix element, mod squared and summed over spins and polarisations, takes the form of a delta comb \cite{Heinzl:2010vg, Heinzl:2009zd},
\begin{gather}\label{SFORM}
	\sum_{\substack{ \text{spins} \\ \text{pols}}}\frac{|S_{fi}^2|}{VT} = \sum\limits_{n>n_0^{\rm nc}} \delta^4( l'+ n k=q + q')  \mathcal{J}_n.
\end{gather}
Consider f\/irst the delta functions (we consider the amplitudes $\mathcal{J}_n$ in a moment). It is the quasi momenta of the particles which are conserved by the delta functions, since
\[
	l'_\mu \equiv k'_\mu + \frac{2a_0^2m^2\sin^2(k\wp k')}{ k\sp k'}k_\mu , \qquad q_\mu \equiv p_\mu + \frac{a_0^2m^2}{ 2k\sp p}k_\mu , \qquad q'_\mu \equiv p'_\mu + \frac{a_0^2m^2}{ 2k\sp p'}k_\mu ,
\]
just as in (\ref{PHOTON.QUASI}) and (\ref{QUASIMOM}) respectively. Thus, our delta comb corresponds to a sum of processes $\gamma'_*  +n\gamma_{\scriptscriptstyle L} \to e^+_* + e^-_*$, in which a dressed probe absorbs $n$ laser photons to produce a heavy pair. Squaring the quasi momentum relations, we obtain the threshold for this process. In QED, the threshold is $n>n_0 = 2m^2_*/k\sp k'$, as discussed in Section~\ref{REVSECT1}, whereas here we f\/ind the threshold number of photons
\begin{gather}\label{REDUCTION}
	n_0^{\rm nc} \equiv \frac{2m_*^2 - {l'}^2/2}{k\sp k'} = \frac{2m^2\left(1+a_0^2 - \Delta^2\right)}{k\sp k'} .
\end{gather}
Noncommutative ef\/fects clearly reduce the threshold for pair production, as $n_0^{\rm nc}< n_0$, and the reduction is proportional to the ef\/fective photon mass squared. Therefore, as was found for Schwinger production in Section~\ref{SCHWING}, the presence of the noncommutative photon reduces the pair production threshold.  This expression, together with our earlier investigation of the photon, makes clear the reason for the reduction. Like the fermions, the noncommutative photon carries additional energy in the form of its dressing, or mass. Hence, less energy -- an amount proportional to the photon mass squared~-- is required from the laser in order to create the pair.

Integrating over the outgoing particle momenta, the total pair production probability  is
\begin{gather}
	W = \frac{e^2m^2 n_\gamma}{32\pi^2 l'_0}\sum\limits_{n>n_0^{\rm nc}}  \int_0^{2\pi} \ud\phi \int_1^{u_{n}} \frac{\ud u}{u\sqrt{u(u-1)}} \nonumber\\
 \phantom{W=}{} \times\bigg[ 2J^2_n + \bigg(a_0^2 - \frac{\Delta^2}{u}\bigg)(2u-1)\left(J^2_{n+1} +  J^2_{n-1} -2 J^2_n\right)\bigg] ,\label{eq:Production_rate}
\end{gather}
where the expressions in large square brackets are the amplitudes $\mathcal{J}_n$ introduced in (\ref{SFORM}), $u$ is the kinematic invariant def\/ined in (\ref{u-def}) and $\phi$ is, in the lab frame, the angle transverse (azimuthal) to the beam.  The argument $z$ of the Bessel functions $J_n$ is the root of
\begin{gather*}
	z^2 \equiv \frac{4u\left(u_{n} - u\right)}{u_1^2\left(1+a_0^2\right)}\left[a_0^2-\frac{\Delta^2}{u}\right] , \qquad u_{n} \equiv \frac{(nk+l')^2}{4m_*^2}= \frac{n}{n_0} + \frac{\Delta^2}{1+a_0^2} .
\end{gather*}
Despite the appealing similarity of (\ref{eq:Production_rate}) to the commutative expression~-- equation (11) of \cite{Narozhnyi:1964}~--  the various arguments are quite complex so, before continuing, let us give a check of our result. Consider the high energy limit in which the probe photon is energetic enough to produce pairs upon interacting with a single background photon, i.e.\ the threshold $n_0^{\rm nc}$ is such that the $n=1$ channel in (\ref{eq:Production_rate}) is open. We also go to low intensity, dialling down the strength of the background so that the probe is incident upon only a few background photons. In other words, we treat the background perturbatively, expanding in the intensity parameter~$a_0$. This should reproduce the high energy Breit--Wheeler process in vacuum, with $a_0^2$ giving a f\/lux factor. We indeed f\/ind that the lowest order contribution comes from
\[
	\frac{\mathcal{J}_1}{a_0^2} \to (2u-1)\bigg( 1-\frac{\sin^2(k\wp k')}{u}\bigg).
\]
Our result therefore correctly recovers the perturbative results described by (\ref{pp_amp}), in terms of undressed particles, i.e.\ Fig.~\ref{NC.3PT} reproduces Fig.~\ref{BW.FIG}. We therefore see, from this limit, that our dressed photon correctly includes the triple photon vertex.

Returning to (\ref{eq:Production_rate}), noncommutative ef\/fects enter the amplitudes $\mathcal{J}_n$ in two ways. First through the explicit apperance of $\Delta^2$, which introduces a dependence on $\phi$, the azimuthal, or transverse angle, because $\Delta$ sees $\mathbf{k}'_\perp$. It follows, since $\theta^{+-}=0$, that a head on collision between the probe and the laser is blind to noncommutative ef\/fects. On the other hand, the total rate is sensitive to, and varies with, the transverse collision angle. Noncommutative ef\/fects also appear in $z$, and def\/ine through this the kinematic ranges $u\in (1,u_{n})$, which are larger than the QED ranges because of the additional energy contribution from $\Delta$. Observe that since $\Delta = a_0\sin(k\wp k')$, all deviations from QED are actually combined `background + noncommutative' ef\/fects, stemming from the behaviour of the noncommutative photon.

It is possible to identify the origins of both types of corrections appearing here. The kinematic changes, including the threshold reduction and changes implicit in $z$, are due to the quasi momentum and ef\/fective mass given to the photon. The {\it explicit} presence of $\Delta^2$ in the amplitudes, though, is due to the polarisation transformation of the photon, which we recall from Section~\ref{PHOTON.SECT} came from a local Lorentz transformation. It is pleasing that a trace of this fairly unique ef\/fect remains in our cross sections (compare \cite{Chatillon:2006rn}).

Currently, optical lasers are the source of the most intense background (near) null f\/ields. While optical lasers are of a low energy, their high intensity compensates for this, to some extent, and the resulting bounds on the noncommutativity parameter which can be obtained from laser based experiments are discussed in \cite{Heinzl:2009zd}.  We give a brief indication of the scales involved. In order to reduce the threshold for stimulated pair creation by one photon, one would require, from (\ref{REDUCTION})
\[
	\frac{l'^2}{2k\sp k'} =\mathcal{O}(1) \implies a_0^2 \frac{\omega\omega'}{\Lambda^2} = \mathcal{O}(1),
\]
where we work to f\/irst order in $\theta$ in the second relation, and $\Lambda$ is the noncommutative energy scale. Hence, even at the maximum allowed optical intensity\footnote{Above this, Schwinger production becomes prevalent and the background is destabilised as photons are transformed into pairs.} of $a_0^2\sim 10^{12}$, extremely high probe photon energies are required to probe sensible noncommutative scales. Nevertheless, through~$\Delta$, increased intensity increases the size of noncommutative ef\/fects, and in bounds such as the above can compensate, partially, for the low energy of the background.  We comment here that backgrounds with compact support provide the scattering amplitudes with a rich structure in QED \cite{Hebenstreit:2009km, Heinzl:2009nd, Mackenroth:2010jk}. Additional noncommutative ef\/fects will also be observed when the background has compact support (in $k\sp x$, as a f\/irst step), resulting from the shifted background seen by the particles and the corrections this introduces into the quasi momenta and ef\/fective mass.

\subsection{Nonlinear Compton scattering}

The crossed process of stimulated pair production is nonlinear Compton scattering, as discussed for strong f\/ield QED in Section~\ref{REVSECT1},  and is forbidden to occur in vacuum, in both QED and NCQED, by momentum conservation. Thus, all deviations from QED are again combined background and noncommutative ef\/fects parameterised by $\Delta =a_0\sin(k\wp k')$, and originate in both the photon quasi momentum and polarisation. The emission rates are expressed in terms of the invariant ${\mathsf x}=k\sp k'/k\sp p'$ and again, the transverse angle, which appears through $k \wedge k'$. Observe that $k\cdot k'$ sees only $k'^+=2k'_-$, while $k\wp k'$ sees only ${\mathbf k}'_\LCperp$. In this case, though, $k'$ is the momentum of the produced photon, i.e.\ an outgoing variable. Consequently, the emission rates display a richer dependence on the scattering angles.  In particular, there is a dependence on the transverse angle at which the photon scatters, to which the QED process is blind, and which agrees qualitatively with the angular distributions found in \cite{Godfrey:2001yy, Godfrey:2001sc} for  Compton scattering in vacuum. Details may be found in \cite{Heinzl:2009zd}. There are also some more novel ef\/fects of noncommutativity, in analogy to the pair production threshold reduction. Let us conclude with an example.

 In QED, the allowed kinematic range of ${\mathsf x}$, which is closely related to the scattered photon frequency in the lab frame, is ${\mathsf x}\in(0,y_n)$ in the $n^\text{th}$ channel, where $y_n=2n k\sp p/m_*^2$. Recall from Section~\ref{REVSECT1}  that the presence of $m_*$ in this expression leads to the redshift of the Compton edge. Now, noncommutative ef\/fects shift these intervals to
\[
	{\mathsf x} >  \frac{a_0^2(k\sp\theta\sp k')^2}{(1+a_0^2)y_n},\qquad {\mathsf x} < y_n - \frac{a_0^2(k\sp\theta\sp k')^2(y_n+1)}{(1+a_0^2)y_n},
\]
working to lowest order in $\theta^{\mu\nu}$. Thus, the lower bound is blueshifted, while the upper bound, which for $n=1$ corresponds to the nonlinear Compton edge, is redshifted by noncommutative ef\/fects. The reason for the blueshift is that the produced photon is dressed, and has an ef\/fective mass, so it's `frequency' cannot be set to zero. Similarly, the nonlinear edge is redshifted relative to strong f\/ield QED because some of the energy transferred to the scattered particles goes into generating the dressing, or mass, of the photon.

\section{Gauge invariant variables in NCQED}\label{GAUGE.SECT}

Finally, we propose a new method for constructing gauge invariant variables in NCQED. Our motivation for doing so is that because the f\/ield strength $F_{\mu\nu}$ is not star-gauge invariant other than when constant, see (\ref{F.TRANS}), it remains to be seen whether our background f\/ields contain the same physics as the gauge invariant QED f\/ield strengths we have associated them with.

While local, gauge invariant operators can be def\/ined in momentum space using Wilson lines~\cite{Gross:2000ba, Das:2000md}, no local invariants can be constructed from the f\/ield strength alone in conf\/iguration space. This is because the equivalent of the Yang--Mills matrix trace (with which one can construct local operators like $\Tr\, F^2$) is an integral $\ud^4x$.  We mention that similar results hold in the standard model, but this is perhaps not widely realised: the physical photon is the transverse part of the gauge f\/ield, after all, and therefore nonlocal. Also, charges in gauge theories are nonlocal, even in QED, as explicitly stated by Gauss' law \cite{Lavelle:1995ty}. Given this, we now propose how one may def\/ine the physical, gauge invariant f\/ields in NCQED. The essential idea originates with Dirac \cite{Dirac:1955uv} and has become very well developed~-- for reviews see \cite{Lavelle:1995ty, Bagan:1999jf, Bagan:1999jk}. It also generalises easily to NCQED, as we now describe.

Begin by choosing some gauge f\/ixing condition, say $\chi=0$, and consider solving
\[
	\chi\big[A^{h[A]}\big] = 0,
\]
for $h$, {\it as a function} of $A$. This gives $h$ as the f\/ield dependent star-gauge transformation which takes an arbitrary f\/ield into the chosen gauge slice. We could of course begin at any other point~$A^U$ on the orbit of $A$, and solve for $h[A^U]$. Now, provided we have a good gauge f\/ixing condition, i.e.\ one which intercepts each gauge orbit once and only once, we can equate in the gauge slice,
\[
	\chi\big[A^{U\star h[A^U]}\big] =  \chi\big[A^{h[A]}\big] ,
\]
which implies that under gauge transformations, the object $h$ transforms as
\begin{gather}\label{h-trans}
	h[A^U] = U^{-1}\star  h[A] .
\end{gather}
This transformation property is the important point, while the gauge condition is, here at least, only relevant as a means of constructing an object which behaves like (\ref{h-trans}). From this, one def\/ines a new potential $\mathcal{A}$,
\begin{gather} \label{35}
	\mathcal{A}_\mu := h^{-1} \star A_\mu \star h +\frac{1}{ie} h^{-1}\star\partial_\mu h ,
\end{gather}
which is star-gauge invariant as is easily checked using using (\ref{A.trans}) and (\ref{h-trans}). Calculating the f\/ield strength in the usual way, one naturally f\/inds
\[
	\mathcal{F}_{\mu\nu} := h^{-1}\star F_{\mu\nu}\star  h,
\]
which is also star-gauge invariant, unlike $F_{\mu\nu}$. Thus, we have an invariant potential and f\/ield strength. Let us give some examples, starting in standard, i.e.\ commutative QED. Consider constructing the transformation into Landau gauge, i.e.\ solving $\partial^\mu(A_\mu^h)=0$ for $h$ as a function of $A_\mu$. This is straightforward, and one obtains
\begin{gather}\label{Landau}
	h[A]= \exp\bigg[ -ie \frac{\partial\sp A}{\partial^2}\bigg]\implies \mathcal{A}_\mu = A_\mu - \frac{1}{\partial^2}\partial_\mu\partial\sp A ,
\end{gather}
which indeed yields the physical photon as the transverse (in the covariant sense) component of the gauge f\/ield\footnote{We work covariantly in this section, in line with the rest of the paper, and so the operator $1/\partial^2$ should be def\/ined by an appropriate $i\epsilon$ prescription. Alternatively, one can work with the Coulomb gauge dressing, replacing $\mu \to j$, spatial, everywhere in this section.}.  The same calculation in NCQED, i.e.\ solving the Landau gauge condition for $h = \exp_\star [ie\alpha(x)]$ is somewhat harder, and akin to the ordinary Yang--Mills calculation. It can of course be tackled perturbatively, in the coupling or in $\theta^{\mu\nu}$. Here we give the f\/irst order noncommutative correction. Writing $\beta\equiv -\partial\sp A/\partial^2$, the QED result, one f\/inds from solving $\partial\sp A^h=0$ in NCQED that
\begin{gather}\label{AB}
	\alpha = \beta - e\frac{\partial_\rho}{\partial^2} \bigg(\partial\beta\sp \theta\sp \partial A_\rho + \frac{1}{2}\partial\beta\sp \theta\sp \partial\partial_\rho\beta\bigg) .
\end{gather}
From this, the gauge invariant f\/ield (\ref{35}) becomes
\begin{gather}\label{AA}
	\mathcal{A}_\mu = A_\mu -\frac{1}{\partial^2}\partial_\mu\partial\sp A -e\partial_\sigma\bigg(\frac{\partial \sp A}{\partial^2}\bigg)\theta^{\sigma\rho}\partial_\rho A_\mu + \frac{e}{2}\partial_\sigma\bigg(\frac{\partial \sp A}{\partial^2}\bigg)\theta^{\sigma\rho}\partial_\rho\partial_\mu\bigg(\frac{\partial \sp A}{\partial^2}\bigg) .
\end{gather}
Thus, we can straightforwardly build gauge invariant f\/ields in NCQED, at least perturbatively. The similarity of (\ref{AB}) and (\ref{AA}) to their Yang--Mills counterparts is very appealing, and suggests more progress in constructing, and using, these gauge invariant f\/ields could be very quickly made.  Let us f\/inish, though, with an example relevant to our strong f\/ield studies, and for which the gauge invariant f\/ields can be computed exactly.

\subsubsection*{Plane waves}
The above simplif\/ies greatly for plane waves, which play the key role of our background f\/ields in this paper. Assume that for a plane wave gauge f\/ield $A(k\sp x)$, the transformation $h$ we are seeking also depends only on $k\sp x$ (through the f\/ield), so that everything is univariate. All star products then become ordinary products, and $A^h$ takes precisely the same form as in QED, i.e.~(\ref{Landau}). Let us be clear: for a general $A_\mu$, the dressed f\/ield $\mathcal{A}_\mu$ is gauge invariant, but constructing it is nontrivial. When $A_\mu$ is a plane wave, however, the transformation to Landau gauge can be written down explicitly, and so we can write down the gauge invariant part of $A$. It is, for this example, the same as the general case in QED, i.e.\ the transverse part. In fact, since our plane wave backgrounds $a_\mu(k\sp x)$ already obey the Landau gauge condition $k\sp a'=0$, we have
\begin{gather}\label{done}
	\mathcal{A} = a_\mu(k\sp x),\qquad \mathcal{F}_{\mu\nu} = k_\mu a'_\nu(k\sp x) - k_\nu a'_\mu(k\sp x).
\end{gather}
We stress these are gauge invariant, by construction. Consequently, the gauge invariant part  of the f\/ield strength matches the commutative f\/ield strength, and the physical content of a plane wave in NCQED is the same as in QED, which, {\it a posteriori}, justif\/ies their use as our background f\/ields.

It is a nice result that one can construct the gauge invariant plane waves (\ref{done}) exactly in NCQED, but the construction of the gauge invariant variables must in general be tackled perturbatively. This brings us to the question of whether such variables will generally exist in NCQED {\it nonperturbatively}, or whether there is a Gribov-like obstruction as in QCD \cite{Ilderton:2007qy}. There, Gribov copies prevent the construction of asymptotic, gauge invariant colour charged variables, which are therefore conf\/ined, but the details depend sensitively on boundary conditions and the topology of the conf\/iguration space~-- for example, see \cite{Ilderton:2010tf} and references therein for how to circumvent the Gribov problem in the electroweak sector, using the QED analogue of (\ref{35}), and \cite{Langfeld:2008tq} for related ideas. We believe these are very interesting questions to explore, which are clearly related to the question of whether bound state, photoball, solutions exist. One can begin by asking, does the $U(1)$ star-gauge symmetry allow for Gribov copies?  It would also be very interesting to see where the Seiberg--Witten map f\/its into this approach \cite{Seiberg:1999vs}.

\section{Conclusions}\label{CONCS.SECT}

\looseness=1
We have considered both classical and quantum aspects of noncommutative QED in a background null f\/ield. We began with the classical Dirac and Maxwell equations,  from which we constructed the propagator of the quantum electron and photon f\/ields in the Furry picture. The poles of the electron propagator conf\/irm that the electron mass shift (which leads to the blueshift of the pair production threshold and red shift of the nonlinear Compton edge) remains unchanged in NCQED, to a good approximation, while the photon dispersion relation and polarisation are signif\/icantly modif\/ied by a combined background and noncommutative ef\/fect. For light-like noncommutativity no restriction on the existence of solutions to the Maxwell equations was found, unlike the case of time-like noncommutativity, see~\cite{Mariz:2006kp}. We also saw evidence for the extended nature of the particles in noncommutative f\/ield theo\-ries, through the advanced and retarded background f\/ields seen by the electron and \mbox{photon}.

We then considered strong f\/ield QED processes using our Furry picture, or dressed, electrons and photons. We found that the noncommutative photon generated all noncommutative corrections, while the dependence of scattering amplitudes on fermions matched that known from QED. The reason for this is that the major ef\/fect of the strong background on fermions, i.e.\ the acquisition of an ef\/fective mass, is unchanged in NCQED. This is reinforced by known results that (generalised) Schwinger pair production in constant f\/ields is unaf\/fected by noncommutative corrections at tree level. At one loop, though, a dressed virtual photon enters and gives the f\/irst correction. This involves, again, a reduction in the pair production threshold, just as we saw a~real photon reduces the threshold in stimulated pair production.

\looseness=1
We note that, experimentally, laser-particle collisions can only place very weak bounds on the noncommutativity scale. The reason is that despite their very high intensities, optical lasers are low energy, and the intensity can only partially compensate for this when probing the high energy noncommutative scale. (High energy, for example gamma ray, lasers are still something of a  pipe dream.) Nevertheless, the noncommutative corrections to strong f\/ield processes can take rather novel forms, as discussed above, and go  beyond the commonly encountered angular dependencies of the cross section.  An important next step is to improve the background f\/ield models employed, especially if one wishes to consider experimental signatures. Retaining the form of a null f\/ield, one can consider f\/inite duration (in $k \sp x$), i.e.\ a laser `pulse'. As shown in QED~\cite{Heinzl:2010vg}, even this simple step alters the form of the cross section considerably. As remarked above, general plane wave backgrounds are also easily accommodated in the NCQED calculations.

Finally, we discussed a new approach to constructing gauge invariant variables in NCQED. Being gauge invariant versions of the Lagrangian f\/ields, these are naturally def\/ined in conf\/iguration space and so are highly nonlocal.  We used this method to conf\/irm that the gauge invariant f\/ield strength for a plane wave background matches the QED f\/ield strength. It is worth pursuing solutions of the noncommutative Maxwell equations in the interest of establishing whether photoball solutions exist, which in turn is connected to the ideas discussed in Section~\ref{GAUGE.SECT}. Since we have seen that the solution for a probe in a plane wave background describes a collection of dressed photons, it may be that the inclusion of a background f\/ield assists in this very interesting problem.

\subsection*{Acknowledgements}

A.I.\ thanks Thomas Heinzl, Martin Lavelle and Christian S\"amann for useful discussions.  M.M.\ and A.I.\ are supported by the European Research Council under Contract No.~204059-QPQV, and the Swedish Research Council under Contract No.~2007-4422. The Feynman diagrams in this article were created using Jaxodraw \cite{Binosi:2003yf, Binosi:2008ig}.

\newpage

\pdfbookmark[1]{References}{ref}
\LastPageEnding


\begin{thebibliography}{99}

\footnotesize\itemsep=0pt


\bibitem{Heinzl:2008an}
Heinzl  T., Ilderton A.,
Exploring high-intensity QED at ELI,
\href{http://dx.doi.org/10.1140/epjd/e2009-00113-x}{{\it Eur.\ Phys.\ J.   D}} {\bf 55} (2009), 359--364,
\href{http://arxiv.org/abs/0811.1960}{arXiv:0811.1960}.


\bibitem{Marklund:2008gj}
 Marklund  M., Lundin J.,
  Quantum vacuum experiments using high intensity lasers,
\href{http://dx.doi.org/10.1140/epjd/e2009-00169-6}{{\it Eur.\ Phys.\ J.~D}} {\bf 55} (2009), 319--326,
\href{http://arxiv.org/abs/0812.3087}{arXiv:0812.3087}.


\bibitem{Landi:1997sh}
Landi  G.,
  An introduction to noncommutative spaces and their geometries, {\it Lecture Notes in Physics. New Series m: Monographs}, Vol.~51, Springer-Verlag, Berlin, 1997,
  \href{http://arxiv.org/abs/hep-th/9701078}{hep-th/9701078}.


\bibitem{Douglas:2001ba}
 Douglas  M.R., Nekrasov N.A.,
Noncommutative f\/ield theory,
\href{http://dx.doi.org/10.1103/RevModPhys.73.977}{{\it Rev. Modern Phys.}}  {\bf 73} (2001), 977--1029,
   \mbox{\href{http://arxiv.org/abs/hep-th/0106048}{hep-th/0106048}}.

\bibitem{Szabo:2001kg}
 Szabo R.J.,
  Quantum f\/ield theory on noncommutative spaces,
\href{http://dx.doi.org/10.1016/S0370-1573(03)00059-0}{{\it Phys.\ Rep.}}  {\bf 378} (2003), 207--299,
 \mbox{\href{http://arxiv.org/abs/hep-th/0109162}{hep-th/0109162}}.

\bibitem{Szabo:2004ic}
Szabo R.J.,
Magnetic backgrounds and noncommutative f\/ield theory,
\href{http://dx.doi.org/10.1142/S0217751X04018099}{{\it Internat. J. Modern Phys. A}} {\bf 19} (2004), 1837--1861,
  \href{http://arxiv.org/abs/physics/0401142}{physics/0401142}.

\bibitem{Carroll:2001ws}
Carroll S.M., Harvey J.A., Kosteleck\'y V.A., Lane C.D., Okamoto T.,
Noncommutative f\/ield theory and Lorentz violation,
\href{http://dx.doi.org/10.1103/PhysRevLett.87.141601}{{\it Phys.\ Rev.\ Lett.}}  {\bf 87} (2001), 141601, 4~pages,
  \href{http://arxiv.org/abs/hep-th/0105082}{hep-th/0105082}.

\bibitem{Szabo:2009tn}
 Szabo R.J.,
  Quantum gravity, f\/ield theory and signatures of noncommutative spacetime,
 \href{http://arxiv.org/abs/0906.2913}{arXiv:0906.2913}.

\bibitem{Kouveliotou:1998ze}
 Kouveliotou  C. et al.,
  An $X$-ray pulsar with a superstrong magnetic f\/ield in the soft $\gamma$-ray repeater SGR 1806-20,
\href{http://dx.doi.org/10.1038/30410}{{\it Nature}} {\bf 393} (1998), 235--237.

\bibitem{Palmer:2005mi}
Palmer  D.M. et al.,
A giant $\gamma$-ray flare from the magnetar SGR 1806-20,
\href{http://dx.doi.org/10.1038/nature03525}{{\it Nature}} {\bf 434} (2005), 1107--1109,
 \href{http://arxiv.org/abs/astro-ph/0503030}{astro-ph/0503030}.

\bibitem{mag3}
Baring M.G., Harding A.K.,
Photon splitting and pair creation in highly magnetized pulsars,
\href{http://dx.doi.org/10.1086/318390}{{\it Astrophys. J.}} {\bf 547} (2001), 929--948,
\href{http://arxiv.org/abs/astro-ph/0010400}{astro-ph/0010400}.


\bibitem{Aharony:2000gz}
 Aharony O., Gomis J., Mehen T.,
On theories with light-like noncommutativity,
\href{http://dx.doi.org/10.1088/1126-6708/2000/09/023}{{\it J. High Energy Phys.}} {\bf 2000} (2000), no.~9,  023, 15~pages,
\href{http://arxiv.org/abs/hep-th/0006236}{hep-th/0006236}.


\bibitem{Gomis:2000zz}
 Gomis J., Mehen T.,
  Space-time noncommutative f\/ield theories and unitarity,
\href{http://dx.doi.org/10.1016/S0550-3213(00)00525-3}{{\it Nuclear Phys.~B}} {\bf 591} (2000), 265--276,
\href{http://arxiv.org/abs/hep-th/0005129}{hep-th/0005129}.

\bibitem{Bahns:2002vm}
Bahns  D., Doplicher S., Fredenhagen K., Piacitelli G.,
  On the unitarity problem in space/time noncommutative theories,
\href{http://dx.doi.org/10.1016/S0370-2693(02)01563-0}{{\it Phys.\ Lett.~B}} {\bf 533} (2002), 178--181,
\href{http://arxiv.org/abs/hep-th/0201222}{hep-th/0201222}.



\bibitem{Heinzl:2009rf}
 Heinzl T., Ilderton A., Marklund M.,
  Noncommutativity and the lightfront,
  \href{http://dx.doi.org/10.1016/j.nuclphysbps.2010.02.021}{{\it Nuclear Phys. Proc. Suppl.}} {\bf 199} (2010), 153--159,
\href{http://arxiv.org/abs/0908.2917}{arXiv:0908.2917}.

\bibitem{Furry:1951zz}
 Furry W.H.,
  On bound states and scattering in positron theory,
\href{http://dx.doi.org/10.1103/PhysRev.81.115}{{\it Phys. Rev.}}  {\bf 81} (1951), 115--124.

\bibitem{MoortgatPick:2009zz}
 Moortgat-Pick G.,
  The Furry picure,
\href{http://dx.doi.org/10.1088/1742-6596/198/1/012002}{{\it J.\ Phys.\ Conf.\ Ser.}}  {\bf 198} (2009), 012002, 7~pages.

\bibitem{Heinzl:2008rh}
 Heinzl T., Ilderton A.,
  A Lorentz and gauge invariant measure of laser intensity,
\href{http://dx.doi.org/10.1016/j.optcom.2009.01.051}{{\it Opt.  Comm.}}  {\bf 282} (2009), 1879--1883,
\href{http://arxiv.org/abs/0807.1841}{arXiv:0807.1841}.

\bibitem{McDonald}
McDonald K.T., Proposal for experimental studies of nonlinear quantum electrodynamics, available at \url{http://www.hep.princeton.edu/~mcdonald/e144/prop.pdf}.


\bibitem{Nikishov:1963}%
 Nikishov A.I., Ritus V.I.,
  Quantum processes in the f\/ield of a plane electromagnetic wave and in a constant f\/ield.~I,
{\it Zh.\ Eksper. Teoret. Fiz.} {\bf 46} (1963), 776--796 (English transl.: {\it Sov.\ Phys.\ JETP} \textbf{19}  (1964),  529--541).

\bibitem{Nikishov:1964a}%
 Nikishov A.I., Ritus V.I.,
  Quantum processes in the f\/ield of a plane electromagnetic wave and in a constant f\/ield.~II,
{\it Zh.\ Eksper. Teoret. Fiz.} {\bf 46} (1964), 1768--1781 (English transl.: {\it Sov.\ Phys.\ JETP} \textbf{19} (1964), 1191--1199).

\bibitem{Narozhnyi:1964}
Narozhnyi N.B., Nikishov A.I., Ritus V.I.,
  Quantum processes in the f\/ield of a circularly polarized electromagnetic wave,
{\it Zh.\ Eksper. Teoret. Fiz.}  {\bf 47} (1964), 930--940  (English transl.: {\it Sov.\ Phys.\ JETP} \textbf{20} (1965), 622--629).

\bibitem{Harvey:2009ry}
 Harvey C., Heinzl T., Ilderton A.,
  Signatures of high-intensity Compton scattering,
\href{http://dx.doi.org/10.1103/PhysRevA.79.063407}{{\it Phys.\ Rev.~A}} {\bf 79} (2009), 063407, 17~pages,
\href{http://arxiv.org/abs/0903.4151}{arXiv:0903.4151}.

\bibitem{Heinzl:2010vg}
 Heinzl T., Ilderton A., Marklund M.,
  Finite size ef\/fects in stimulated laser pair production,
\href{http://arxiv.org/abs/1002.4018}{arXiv:1002.4018}.

\bibitem{Toll:1952}
 Toll J., The dispersion relation for light and its application to problems involving electron pairs,
 PhD Thesis, Princeton, 1952.

\bibitem{Heinzl:2006xc}
 Heinzl T., Liesfeld B., Amthor K.U., Schwoerer H., Sauerbrey R., Wipf A.,
 On the observation of vacuum birefringence,
\href{http://dx.doi.org/10.1016/j.optcom.2006.06.053}{{\it Opt.\ Comm.}}  {\bf 267} (2006), 318--321,
\href{http://arxiv.org/abs/hep-ph/0601076}{hep-ph/0601076}.

\bibitem{Abel:2006wj}
Abel  S.A., Jaeckel J., Khoze V.V., Ringwald A.,
  Vacuum birefringence as a probe of Planck scale noncommutativity,
\href{http://dx.doi.org/10.1088/1126-6708/2006/09/074}{{\it J. High Energy Phys.}} {\bf 2006} (2006), no.~9, 074, 18~pages,
\href{http://arxiv.org/abs/hep-ph/0607188}{hep-ph/0607188}.


\bibitem{Seiberg:1999vs}
 Seiberg N., Witten E.,
  String theory and noncommutative geometry,
\href{http://dx.doi.org/10.1088/1126-6708/1999/09/032}{{\it J. High Energy Phys.}} {\bf 1999} (1999), no.~9, 032, 93~pages,
\href{http://arxiv.org/abs/hep-th/9908142}{hep-th/9908142}.

\bibitem{Gross:2000ba}
 Gross D.J., Hashimoto A., Itzhaki N.,
  Observables of non-commutative gauge theories,
{\it Adv.\ Theor.\ Math.\ Phys.}  {\bf 4} (2000), 893--928,
\href{http://arxiv.org/abs/hep-th/0008075}{hep-th/0008075}.

\bibitem{Langmann:2001yr}
 Langmann E., Szabo R.J.,
  Teleparallel gravity and dimensional reductions of noncommutative gauge theory,
\href{http://dx.doi.org/10.1103/PhysRevD.64.104019}{{\it Phys.\ Rev.~D}} {\bf 64} (2001), 104019, 15~pages,
\href{http://arxiv.org/abs/hep-th/0105094}{hep-th/0105094}.


\bibitem{Hayakawa:1999yt}
 Hayakawa M.,
  Perturbative analysis on infrared aspects of noncommutative QED on  ${\mathbb R}^4$,
\href{http://dx.doi.org/10.1016/S0370-2693(00)00242-2}{{\it Phys.\ Lett.\  B}} {\bf 478} (2000), 394--400,
\href{http://arxiv.org/abs/hep-th/9912094}{hep-th/9912094}.

\bibitem{Hayakawa:1999zf}
 Hayakawa M.,
  Perturbative analysis on infrared and ultraviolet aspects of noncommutative QED on ${\mathbb R}^4$,
\href{http://arxiv.org/abs/hep-th/9912167}{hep-th/9912167}.

\bibitem{Hewett:2000zp}
 Hewett J.L., Petriello F.J., Rizzo T.G.,
  Signals for non-commutative interactions at linear colliders,
\href{http://dx.doi.org/10.1103/PhysRevD.64.075012}{{\it Phys.\ Rev.\  D}} {\bf 64} (2001), 075012, 23~pages,
  \href{http://arxiv.org/abs/hep-ph/0010354}{hep-ph/0010354}.

\bibitem{Ardalan:2000qk}
 Ardalan F., Sadooghi N.,
  Anomaly and nonplanar diagrams in noncommutative gauge theories,
\href{http://dx.doi.org/10.1142/S0217751X02005694}{{\it Internat. J. Modern Phys. A}} {\bf 17} (2002), 123--144,
   \href{http://arxiv.org/abs/hep-th/0009233}{hep-th/0009233}.

\bibitem{Nakajima:2001uh}
 Nakajima T.,
  Conformal anomalies in noncommutative gauge theories,
\href{http://dx.doi.org/10.1103/PhysRevD.66.085008}{{\it Phys.\ Rev.\  D}} {\bf 66} (2002), 085008, 9~pages,
 \href{http://arxiv.org/abs/hep-th/0108158}{hep-th/0108158}.

\bibitem{Minwalla:1999px}
 Minwalla S., Van Raamsdonk M., Seiberg N.,
 Noncommutative perturbative dynamics,
\href{http://dx.doi.org/10.1088/1126-6708/2000/02/020}{{\it J. High Energy Phys.}} {\bf 2000} (2000), no.~2,  020, 31~pages,
\href{http://arxiv.org/abs/hep-th/9912072}{hep-th/9912072}.

\bibitem{Matusis:2000jf}
 Matusis A., Susskind L., Toumbas N.,
  The IR/UV connection in the non-commutative gauge theories,
\href{http://dx.doi.org/10.1088/1126-6708/2000/12/002}{{\it J.~High Energy Phys.}} {\bf 2000} (2000), no.~12, 002, 18~pages,
\href{http://arxiv.org/abs/hep-th/0002075}{hep-th/0002075}.

\bibitem{Martin:1999aq}
 Mart\'{\i}n C.P., S\'anchez-Ruiz D.,
One-loop UV divergent structure of $U(1)$ Yang--Mills theory on noncommutative ${\mathbb R}^4$,
\href{http://dx.doi.org/10.1103/PhysRevLett.83.476}{{\it Phys.\ Rev.\ Lett.}}  {\bf 83} (1999), 476--479,
\href{http://arxiv.org/abs/hep-th/9903077}{hep-th/9903077}.

\bibitem{Grosse:2004yu}
 Grosse  H., Wulkenhaar R.,
 Renormalisation of $\phi^4$ theory on noncommutative ${\mathbb R}^4$ in the matrix   base,
\href{http://dx.doi.org/10.1007/s00220-004-1285-2}{{\it Comm.\ Math.\ Phys.}}  {\bf 256} (2005), 305--374,
\href{http://arxiv.org/abs/hep-th/0401128}{hep-th/0401128}.

\bibitem{deGoursac:2010zb}
 de Goursac A.,
  On the origin of the harmonic term in noncommutative quantum f\/ield theory,
\href{http://arxiv.org/abs/1003.5788}{arXiv:1003.5788}.

\bibitem{de Goursac:2007gq}
 de Goursac A., Wallet J.-C., Wulkenhaar R.,
  Noncommutative induced gauge theory,
\href{http://dx.doi.org/10.1140/epjc/s10052-007-0335-2}{{\it Eur. Phys. J. C}} {\bf 51} (2007), 977--987,
\href{http://arxiv.org/abs/hep-th/0703075}{hep-th/0703075}.

\bibitem{Grosse:2007dm}
 Grosse  H., Wohlgenannt M.,
  Induced gauge theory on a noncommutative space,
\href{http://dx.doi.org/10.1140/epjc/s10052-007-0369-5}{{\it Eur.\ Phys.~J.~C}} {\bf 52} (2007), 435--450,
\href{http://arxiv.org/abs/hep-th/0703169}{hep-th/0703169}.


\bibitem{deGoursac:2009gh}
 de Goursac A.M.,
  Noncommutative geometry, gauge theory and renormalization, PhD Thesis,
\href{http://arxiv.org/abs/0910.5158}{arXiv:0910.5158}.

\bibitem{Grosse:2006hh}
 Grosse H., Wohlgenannt M.,
  Noncommutative QFT and renormalization,
\href{http://dx.doi.org/10.1088/1742-6596/53/1/050}{{\it J.\ Phys.\ Conf.\ Ser.}}  {\bf 53} (2006), 764--792,
\href{http://arxiv.org/abs/hep-th/0607208}{hep-th/0607208}.

\bibitem{Rivasseau:2007rz}
 Rivasseau  V., Vignes-Tourneret F.,
  Renormalisation of non-commutative f\/ield theories,
\href{http://arxiv.org/abs/hep-th/0702068}{hep-th/0702068}.

\bibitem{Blaschke:2009rb}
 Blaschke D.N., Kronberger E., Rofner A., Schweda M., Sedmik R.I.P., Wohlgenannt M.,
  On the problem of renormalizability in non-commutative gauge f\/ield models -- a critical review,
\href{http://dx.doi.org/10.1002/prop.200900102}{{\it Fortschr. Phys.}} {\bf 58} (2010), 364--372,
\href{http://arxiv.org/abs/0908.0467}{arXiv:0908.0467}.


\bibitem{Fischer:2010zg}
 Fischer  A., Szabo R.J.,
 UV/IR duality in noncommutative quantum f\/ield theory,
\href{http://arxiv.org/abs/1001.3776}{arXiv:1001.3776}.

\bibitem{Langmann:2002cc}
 Langmann E., Szabo R.J.,
  Duality in scalar f\/ield theory on noncommutative phase spaces,
\href{http://dx.doi.org/10.1016/S0370-2693(02)01650-7}{{\it Phys.\ Lett.~B}} {\bf 533} (2002), 168--177,
\href{http://arxiv.org/abs/hep-th/0202039}{hep-th/0202039}.

\bibitem{Blaschke:2008yj}
 Blaschke D.N., Gieres F., Kronberger E., Schweda M., Wohlgenannt M.,
  Translation-invariant models for non-commutative gauge f\/ields,
\href{http://dx.doi.org/10.1088/1751-8113/41/25/252002}{{\it J.\ Phys.\ A: Math. Theor.}}  {\bf 41} (2008), 252002, 7~pages,
\href{http://arxiv.org/abs/0804.1914}{arXiv:0804.1914}.

\bibitem{Schupp:2008fs}
 Schupp  P., You J.,
  UV/IR mixing in noncommutative QED def\/ined by Seiberg--Witten map,
\href{http://dx.doi.org/10.1088/1126-6708/2008/08/107}{{\it J. High Energy Phys.}} {\bf 2008} (2008), no.~8, 107, 10~pages,
  \href{http://arxiv.org/abs/0807.4886}{arXiv:0807.4886}.

\bibitem{Raasakka:2010ev}
 Raasakka  M., Tureanu A.,
  On UV/IR mixing via Seiberg--Witten map for noncommutative QED,
  \href{http://arxiv.org/abs/1002.4531}{arXiv:1002.4531}.

\bibitem{AlvarezGaume:2000bv}
 Alvarez-Gaum\'e  L., Barb\'on J.L.F.,
  Non-linear vacuum phenomena in non-commutative QED,
\href{http://dx.doi.org/10.1142/S0217751X01002759}{{\it Internat. J. Modern Phys. A}} {\bf 16} (2001), 1123--1146,
  \href{http://arxiv.org/abs/hep-th/0006209}{hep-th/0006209}.

\bibitem{Volkov:1935}
 Wolkow D.,
  \"Uber eine Klasse von L\"osungen der Diracschen Gleichung,
\href{http://dx.doi.org/10.1007/BF01331022}{{\it Z.~Phys.}}  {\bf 94} (1935), 250--260.

\bibitem{Barbon:2001bi}
 Barb\'on J.L.F.,
  Introduction to noncommutative f\/ield theory,
in Proceedings of ICTP Summer School in Particle Physics (Trieste, Italy,  June 18 --  July 6, 2001),
Editors A.~Masiero, G.~Senjanovic, A.Yu.~Smirnov, G.~Thompson,
{\it ICTP Lecture Note Series}, Vol.~10, Trieste, 2002, 185--219.

\bibitem{Gerhold:2000ik}
 Gerhold A., Grimstrup J., Grosse H., Popp L., Schweda M., Wulkenhaar R.,
  The energy-momentum tensor on noncommutative spaces: some pedagogical comments,
  \href{http://arxiv.org/abs/hep-th/0012112}{hep-th/0012112}.

\bibitem{Das:2002jd}
 Das  A.K., Frenkel J.,
Energy-momentum tensor in non-commutative gauge theories,
\href{http://dx.doi.org/10.1103/PhysRevD.67.067701}{{\it Phys.\ Rev.   D}} {\bf 67} (2003), 067701, 3~pages,
 \href{http://arxiv.org/abs/hep-th/0212122}{hep-th/0212122}.


\bibitem{Brown:1964zz}
 Brown  L.S., Kibble T.W.B.,
  Interaction of intense laser beams with electrons,
\href{http://dx.doi.org/10.1103/PhysRev.133.A705}{{\it Phys.\ Rev.}}  {\bf 133} (1964), A705--A719.

\bibitem{Reiss:Greens}
Reiss H.R., Eberly J.H.,
Green's function in intense-f\/ield electrodynamics,
\href{http://dx.doi.org/10.1103/PhysRev.151.1058}{{\it Phys.\ Rev.}}  {\bf 151} (1966), 1058--1066.


\bibitem{Dymarsky:2001xg}
 Dymarsky A.Y.,
 Noncommutative f\/ield theory in formalism of f\/irst quantization,
\href{http://dx.doi.org/10.1016/S0370-2693(02)01166-8}{{\it Phys.\ Lett.~B}} {\bf 527} (2002), 125--130,
  \href{http://arxiv.org/abs/hep-th/0104250}{hep-th/0104250}.

\bibitem{Coleman:1977ps}
 Coleman S.,
  Non-Abelian plane waves,
\href{http://dx.doi.org/10.1016/0370-2693(77)90344-6}{{\it Phys.\ Lett.\  B}} {\bf 70} (1977), 59--60.


\bibitem{Mariz:2006kp}
 Mariz T., Nascimento J.R., Rivelles V.O.,
  Dispersion relations in noncommutative theories,
\href{http://dx.doi.org/10.1103/PhysRevD.75.025020}{{\it Phys.\ Rev.~D}} {\bf 75} (2007), 025020, 7~pages,
   \href{http://arxiv.org/abs/hep-th/0609132}{hep-th/0609132}.

\bibitem{Guralnik:2001ax}
 Guralnik Z., Jackiw R., Pi S.Y., Polychronakos A.P.,
  Testing non-commutative QED, constructing non-commutative MHD,
\href{http://dx.doi.org/10.1016/S0370-2693(01)00986-8}{{\it Phys.\ Lett.~B}} {\bf 517} (2001), 450--456,
   \href{http://arxiv.org/abs/hep-th/0106044}{hep-th/0106044}.

\bibitem{Chatillon:2006rn}
 Chatillon  N., Pinzul A.,
  Light propagation in a background f\/ield for time-space non-commutativity and axionic non-commutative QED,
\href{http://dx.doi.org/10.1016/j.nuclphysb.2006.12.003}{{\it Nuclear Phys.~B}} {\bf 764} (2007), 95--108,
  \href{http://arxiv.org/abs/hep-ph/0607243}{hep-ph/0607243}.

\bibitem{Abbott:1981ke}
 Abbott L.F.,
  Introduction to the background f\/ield method,
{\it Acta Phys.\ Polon.~B} {\bf 13} (1982), 33--50.

\bibitem{Fatollahi:2005ri}
 Fatollahi A.H., Jafari A.,
  On the bound states of photons in noncommutative $U(1)$ gauge theory,
\href{http://dx.doi.org/10.1140/epjc/s2005-02465-8}{{\it Eur.\ Phys.~J.~C}} {\bf 46} (2006), 235--245,
   \href{http://arxiv.org/abs/hep-th/0503078}{hep-th/0503078}.

\bibitem{Volkholz:2005mp}
 Volkholz J., Bietenholz W., Nishimura J., Susaki Y.,
  The scaling of QED in a non-commutative space-time,
{\it PoS LAT2005} (2005), 264, 6~pages,
  \href{http://arxiv.org/abs/hep-lat/0509146}{hep-lat/0509146}.

\bibitem{Gomis:2000pf}
 Gomis J., Landsteiner K., Lopez E.,
 Non-relativistic non-commutative f\/ield theory and UV/IR mixing,
\href{http://dx.doi.org/10.1103/PhysRevD.62.105006}{{\it Phys.\ Rev.\  D}} {\bf 62} (2000), 105006, 6~pages,
   \href{http://arxiv.org/abs/hep-th/0004115}{hep-th/0004115}.


\bibitem{Baek:2001ty}
 Baek S., Ghosh D.K., He X.G., Hwang W.Y.P.,
  Signatures of non-commutative QED at photon colliders,
\href{http://dx.doi.org/10.1103/PhysRevD.64.056001}{{\it Phys.\ Rev.\  D}} {\bf 64} (2001), 056001, 7~pages,
  \href{http://arxiv.org/abs/hep-ph/0103068}{hep-ph/0103068}.

\bibitem{Godfrey:2001yy}
 Godfrey  S., Doncheski M.A.,
  Signals for non-commutative QED in $e\gamma$ and $\gamma\gamma$ collisions,
\href{http://dx.doi.org/10.1103/PhysRevD.65.015005}{{\it Phys.\ Rev.\  D}} {\bf 65} (2002), 015005, 9~pages,
  \href{http://arxiv.org/abs/hep-ph/0108268}{hep-ph/0108268}.

\bibitem{Godfrey:2001sc}
 Godfrey  S., Doncheski M.A.,
  Signals for non-commutative QED in $e\gamma$ and $\gamma\gamma$ collisions,
in  Proceedings of the APS/DPF/DPB Summer Study on the Future of Particle Physics (Snowmass, 2001), Editor N.~Graf,
2001, P313, 4~pages,
 \href{http://arxiv.org/abs/hep-ph/0111147}{hep-ph/0111147}.


\bibitem{Ohl:2004tn}
 Ohl  T., Reuter J.,
  Testing the noncommutative standard model at a future photon collider,
\href{http://dx.doi.org/10.1103/PhysRevD.70.076007}{{\it Phys.\ Rev.\  D}} {\bf 70} (2004), 076007, 10~pages,
  \href{http://arxiv.org/abs/hep-ph/0406098}{hep-ph/0406098}.

\bibitem{Sauter:1931zz}
 Sauter F.,
  \"Uber das Verhalten eines Elektrons im homogenen elektrischen Feld nach der relativistischen Theorie Diracs,
\href{http://dx.doi.org/10.1007/BF01339461}{{\it Z.\ Phys.}}  {\bf 69}  (1931), 742--764.


\bibitem{Schwinger:1951nm}
 Schwinger J.S.,
  On gauge invariance and vacuum polarization,
\href{http://dx.doi.org/10.1103/PhysRev.82.664}{{\it Phys.\ Rev.}}  {\bf 82} (1951), 664--679.


\bibitem{Chair:2000vb}
 Chair  N., Sheikh-Jabbari M.M.,
  Pair production by a constant external f\/ield in noncommutative QED,
\href{http://dx.doi.org/10.1016/S0370-2693(01)00259-3}{{\it Phys.\ Lett.   B}} {\bf 504} (2001), 141--146,
\href{http://arxiv.org/abs/hep-th/0009037}{hep-th/0009037}.


\bibitem{Riad:2000vy}
 Riad  I.F., Sheikh-Jabbari M.M.,
  Noncommutative QED and anomalous dipole moments,
\href{http://dx.doi.org/10.1088/1126-6708/2000/08/045}{{\it J. High Energy Phys.}} {\bf 2000} (2000), no.~8, 045, 22~pages,
  \href{http://arxiv.org/abs/hep-th/0008132}{hep-th/0008132}.


\bibitem{Heinzl:2009zd}
 Heinzl T., Ilderton A., Marklund M.,
  Laser intensity ef\/fects in noncommutative QED,
  \href{http://dx.doi.org/10.1103/PhysRevD.81.051902}{{\it Phys. Rev. D}} {\bf 81} (2010), 051902, 5~pages,
   \href{http://arxiv.org/abs/0909.0656}{arXiv:0909.0656}.


\bibitem{Hebenstreit:2009km}
 Hebenstreit F., Alkofer R., Dunne G.V., Gies H.,
  Momentum signatures for Schwinger pair production in short laser pulses with sub-cycle structure,
\href{http://dx.doi.org/10.1103/PhysRevLett.102.150404}{{\it Phys.\ Rev.\ Lett.}}  {\bf 102} (2009), 150404, 4~pages,
   \href{http://arxiv.org/abs/0901.2631}{arXiv:0901.2631}.

\bibitem{Heinzl:2009nd}
 Heinzl T., Seipt D., Kampfer B.,
  Beam-shape ef\/fects in nonlinear Compton and Thomson scattering,
\href{http://dx.doi.org/10.1103/PhysRevA.81.022125}{{\it Phys.\ Rev.~A}} {\bf 81} (2010), 022125, 17~pages,
  \href{http://arxiv.org/abs/0911.1622}{arXiv:0911.1622}.

\bibitem{Mackenroth:2010jk}
 Mackenroth F., Di Piazza A., Keitel C.H.,
  Determining the carrier-envelope phase of short intense laser pulses,
  \href{http://arxiv.org/abs/1001.3614}{arXiv:1001.3614}.

\bibitem{Das:2000md}
 Das  S.R., Rey S.J.,
 Open Wilson lines in noncommutative gauge theory and tomography of holographic dual supergravity,
\href{http://dx.doi.org/10.1016/S0550-3213(00)00549-6}{{\it Nuclear Phys.~B}} {\bf 590} (2000), 453--470,
  \href{http://arxiv.org/abs/hep-th/0008042}{hep-th/0008042}.

\bibitem{Lavelle:1995ty}
 Lavelle  M., McMullan D.,
  Constituent quarks from QCD,
\href{http://dx.doi.org/10.1016/S0370-1573(96)00019-1}{{\it  Phys.\ Rep.}}  {\bf 279} (1997), 1--65,
  \href{http://arxiv.org/abs/hep-ph/9509344}{hep-ph/9509344}.

\bibitem{Dirac:1955uv}
 Dirac P.A.M.,
  Gauge-invariant formulation of quantum electrodynamics,
{\it Canad. J. Phys.}  {\bf 33} (1955), 650--660.

\bibitem{Bagan:1999jf}
 Bagan E., Lavelle M., McMullan D.,
  Charges from dressed matter: construction,
\href{http://dx.doi.org/10.1006/aphy.2000.6048}{{\it Ann. Physics}}  {\bf 282} (2000), 471--502,
  \href{http://arxiv.org/abs/hep-ph/9909257}{hep-ph/9909257}.

\bibitem{Bagan:1999jk}
 Bagan E., Lavelle M., McMullan D.,
  Charges from dressed matter: physics and renormalisation,
\href{http://dx.doi.org/10.1006/aphy.2000.6049}{{\it Ann. Physics}}  {\bf 282} (2000), 503--540,
   \href{http://arxiv.org/abs/hep-ph/9909262}{hep-ph/9909262}.

\bibitem{Ilderton:2007qy}
 Ilderton A., Lavelle M., McMullan D.,
  Colour, copies and conf\/inement,
\href{http://dx.doi.org/10.1088/1126-6708/2007/03/044}{{\it J. High Energy Phys.}} {\bf 2007} (2007), no.~3, 044, 27~pages,
  \href{http://arxiv.org/abs/hep-th/0701168}{hep-th/0701168}.

\bibitem{Ilderton:2010tf}
Ilderton A., Lavelle M., McMullan D.,
Symmetry breaking, conformal geometry and gauge invariance,
  \href{http://arxiv.org/abs/1002.1170}{arXiv:1002.1170}.

\bibitem{Langfeld:2008tq}
 Langfeld K., Heinzl T., Ilderton A., Lavelle M., McMullan D.,
  Solution of the Gribov problem from gauge invariance,
 {\it PoS(CONFINEMENT8)} (2008), 047, 5~pages,
  \href{http://arxiv.org/abs/0812.2418}{arXiv:0812.2418}.

\bibitem{Binosi:2003yf}
 Binosi  D., Theussl L.,
  JaxoDraw: a graphical user interface for drawing Feynman diagrams,
\href{http://dx.doi.org/10.1016/j.cpc.2004.05.001}{{\it Comput.\ Phys.\ Comm.}}  {\bf 161} (2004), 76--86,
   \href{http://arxiv.org/abs/hep-ph/0309015}{hep-ph/0309015}.

\bibitem{Binosi:2008ig}
 Binosi D., Collins J., Kaufhold C., Theussl L.,
  JaxoDraw: a graphical user interface for drawing Feynman diagrams. Version 2.0 release notes,
\href{http://dx.doi.org/10.1016/j.cpc.2009.02.020}{{\it Comput.\ Phys.\ Comm.}}  {\bf 180} (2009), 1709--1715,
   \href{http://arxiv.org/abs/0811.4113}{arXiv:0811.4113}.


\end{thebibliography}
\end{document}